\documentclass[acmsmall,natbib=false,screen,nonacm]{acmart} 

\AtBeginDocument{
  }

\RequirePackage[
  datamodel=acmdatamodel,
  style=acmauthoryear,
  ]{biblatex}

\usepackage[hyperfirst=false]{glossaries}
\newacronym{CM}{CM}{Configuration Model}
\newacronym{UBCM}{UBCM}{Undirected Binary Configuration Model}
\newacronym{DBCM}{DBCM}{Directed Binary Configuration Model}
\newacronym{BiCM}{BiCM}{Bipartite Configuration Model}
\newacronym{RBCM}{RBCM}{Reciprocal Binary Configuration Model} 
\newacronym{LRA}{LRA}{Local Rewiring Algorithm}

\begin{document}

\title{Comparative analysis of graph randomization: Tools, methods, pitfalls, and best practices}

\author{Bart De Clerck}
\email{bart.declerck@ugent.be}
\orcid{0000-0002-9718-260X}
\affiliation{%
  \institution{Ghent University, Faculty of Economics and Business Administration}
  \city{Ghent}
  \country{Belgium}
}
\affiliation{
  \institution{Royal Military Academy, Department of Mathematics}
  \city{Brussels}
  \country{Belgium}
}

\author{Filip Van Utterbeeck}
\affiliation{
  \institution{Royal Military Academy, Department of Mathematics}
  \city{Brussels}
  \country{Belgium}
}

\author{Luis E.C. Rocha}
\orcid{0000-0001-9046-8739}
\affiliation{%
  \institution{Ghent University, Faculty of Economics and Business Administration, Dept Economics and Dept Physics and Astronomy}
  \city{Ghent}
  \country{Belgium}
}

\renewcommand{\shortauthors}{De Clerck et al.}

\begin{abstract}
Graph randomization techniques play a crucial role in network analysis, 
allowing researchers to assess the statistical significance of observed network properties and distinguish meaningful patterns from random fluctuations. 
In this survey we provide an overview of the graph randomization methods available in the most popular software tools for network analysis. 
We propose a comparative analysis of popular software tools to highlight their functionalities and limitations. 
Through case studies involving diverse graph types, we demonstrate how different randomization methods can lead to divergent conclusions, 
emphasizing the importance of careful method selection based on the characteristics of the observed network and the research question at hand. 
This survey proposes some guidelines for researchers and practitioners seeking to understand and 
utilize graph randomization techniques effectively in their network analysis projects.
\end{abstract}

\maketitle

\section{Introduction}
An ever increasing number of scientific disciplines discover the power of network science to model and analyze the complex systems they study. 
Some recent examples include the use of network science in archaeology \cite{brughmans_peeples_2023},  
biology \cite{GOSAK2018118,https://doi.org/10.1111/oik.09436,Klaise:2017aa,10.1371/journal.pcbi.1011624}, 
criminology \cite{BRIGHT202150}, 
ecology \cite{https://doi.org/10.1111/2041-210X.13149}, 
economics \cite{Saracco:2015aa,10.1007/978-3-642-28583-7_3},
history \cite{https://doi.org/10.25517/jhnr.v5i1}, 
law \cite{9568271, coupette_corinna_2019_2617115}, 
recommender systems \cite{PhysRevE.99.022306}, 
sociology \cite{rice2015social} etc. 
In all these fields, networks provided a suitable framework to model, 
analyze, and understand the structure, function, and dynamics of the systems under investigation.
In terms of terminology, we follow the convention where a graph refers to a mathematical entity, 
whereas a network refers to a real-world system that can be described by a graph \cite{vitolatora2017, ginestrabianconi2018}.\\

In many of these applications, a metric or topological aspect of a network is used to draw conclusions about the system under investigation. 
Frequently, the value of a metric for an observed network alone does not provide sufficient information to draw a conclusion about the subject being investigated. 
Analyzing the extent to which the observed network is ``unexpected'' or ``atypical'' can genuinely result in new and unique insights. 
Consequently, we often depend on comparing observed measurements or topologies with those generated by a randomization process. 
These comparisons help to differentiate important structural patterns from those that are more likely to happen by chance. 
In order to quantify this phenomenon in a rigorous manner, one can employ null models to construct a sample of graphs that represent the observed system, 
and subsequently analyze the sample using statistical methods. 
Nevertheless, there are numerous methods and tools for creating random graphs, along with several approaches for assessing statistical significance. 
Under certain scenarios, the technique used for generating samples, together with the approach used to assess statistical significance, 
can result in significantly divergent results.\\

\emph{The survey focus}. Oftentimes, researchers and practitioners will use readily available software tools for graph randomization, 
leveraging the advanced functionalities these tools offer. 
However, the effectiveness of such tools can greatly depend on one's familiarity with their underlying assumptions and algorithms. 
It can be challenging to select the most suitable method for specific research questions without a clear understanding of these aspects. 
This review aims to close the gap by providing an overview of graph randomization algorithms used in popular software applications along 
with a comparative analysis that highlights their features and limitations. \\

\emph{Contributions}. By conducting a comparative analysis of graph randomization tools, 
we reveal what different tools have in common and where they differ, 
both in terms of the conceptual approach and the algorithmic implementation. 
By providing case studies, we illustrate how different randomization methods can lead to divergent conclusions, 
emphasizing the importance of careful method selection. 
Highlighting this issue is important for researchers who wish to understand and utilize graph randomization techniques effectively in their network analysis projects. 
Additionally, by emphasizing the importance of fully detailing the randomization process in research papers, 
we aim to improve the transparency and reproducibility of network analysis studies.\\

The rest of the paper is structured as follows. 
In the upcoming sections, we will first concentrate on the conceptual approaches to graph randomization and 
the corresponding algorithmic procedures for creating random graphs. 
The primary focus will be on the features provided by the most widely used software tools, which will be discussed. 
Throughout different case studies, we will illustrate how various randomization procedures might result in contrasting outcomes. 
In addition, we will look at the usage of statistical evaluation in the randomization of graphs. 
Lastly, we will offer guidelines and suggestions for choosing suitable randomization techniques and 
tools that align with particular research questions.

\section{Conceptual Approaches to Graph Randomization}
We will limit our discussion to class of graph models that are static, i.e. not evolving over time, 
that take as input a (list of) feature(s) or constraints, desired to be reproduced in the randomized graphs. 
In the present discussion, we also exclude weighted graphs, due to their limited availability in the tools we will discuss. 
Additionally, we will focus specifically on graph models that used the degree sequence as a constraint. 
The field of graph randomization is vast and encompasses a wide range of methods and models. 
For an extensive survey and a taxonomy of the concepts of random graph modeling we refer to \cite{Drobyshevskiy_2020}.\\

A single graph, defined as $G(V,E)$ is composed of its set of vertices (or nodes), $V$, and its set of edges (or links) $E$. 
We distinguish between directed and undirected graphs, depending on whether an edge from node $i$ to node $j$ is automatically reciprocal.
We also consider bipartite graphs, defined as $G(U,V,E)$. These are composed of two disjoint sets of vertices, $U$ and $V$, 
and the set of edges $E$, where an edge can only connect a node from $U$ to a node from $V$, i.e., $U \cap V = \emptyset$.\\

A random graph is a graph that is created using a stochastic process, where certain characteristics are predetermined, 
but the graph itself is otherwise random \cite{newman2018networks}. 
Often, we like our random graph to possess specific characteristics that are observed in the actual network under investigation. 
For instance, it may be desirable for the degree sequence of the random graph to correspond to one that has been seen. 
A graph ensemble $\mathcal{G}$ refers to a collection of graphs that possess the same set of properties or constraints. 
Each graph in the ensemble is typically created using a randomization method. 
The ensemble is characterized by the probability distribution across this set of graphs, 
which determines the likelihood of selecting or realizing each graph from the ensemble. 
In formal terms, a graph ensemble is a probability space in which each element represents a graph. 
The ensemble is characterized by a probability measure ($P$) that assigns a probability ($P(G)$) to each graph ($G$) in the ensemble. 
The total of the probability of all graphs in the ensemble is equivalent to one, i.e., 
 
\begin{equation}
    \sum_{G \in \mathcal{G}} P(G) = 1
    \label{eq:random:sumprob}
\end{equation}

The primary distinction between randomization approaches lies in the manner in which the constraints are imposed. 
There is a distinction between hard and soft constraints. 
Hard constraints must be satisfied by every randomized graph in the ensemble. 
This ensemble is sometimes referred to as the microcanonical ensemble \cite{ginestrabianconi2018}. 
The randomization procedure must be able to generate all possible graphs that satisfy constraints with equal probability. 
Formally, if we use $G^{*}$ to denote the observed graph, we can express this as follows \cite{squartini2017maximum-entropy}: 
\begin{equation}
  \forall G \in \mathcal{G} : P(G) = \begin{cases}
      \frac{1}{\vert \mathcal{G} \vert} & \text{if } \mathbf{C}(G) = \mathbf{C}(G^{*}); \\ 
      0 & \text{otherwise},
  \end{cases}
\end{equation}
where $\mathbf{C}(G^{*})$ denotes the features of the observed graph imposed on the ensemble.
In the case of soft constraints, the randomization process must be able to generate all possible graphs that, on average, satisfy the constraints.
Formally, we can express this as follows:
\begin{equation}
  \mathbf{C}(\mathcal{G}) = \langle \mathbf{C} \rangle = \sum_{G \in \mathcal{G}} \mathbf{C}(G)P(G) = \mathbf{C}(G^{*}).
\end{equation}
This ensemble is sometimes referred to as the canonical ensemble \cite{ginestrabianconi2018}.
It is important to note that the microcanonical and canonical ensembles are non-equivalent in the thermodynamic limit 
for models of networks with an extensive number of constraints \cite{realstats,PhysRevE.80.045102,PhysRevLett.115.268701}.\\

The canonical ensemble is more suitable for systems affected by measurement errors, incomplete data, 
or random noise due to its ability to withstand mistake in the original data. 
The microcanonical ensemble sets strict limitations and does not permit statistical changes in observed values. 
The selection between microcanonical and canonical methods for selecting network ensembles is a theoretical topic that requires a formal analysis. 
The canonical ensemble is often used due to its mathematical tractability, unbiased nature, 
and ability to manage imperfections in the data \cite{squartini2017maximum-entropy}.

\subsection{Microcanonical ensembles}
\subsubsection{Erdős-Rényi model ${\mathcal{G}(N,L)}$}
The uniform random graph ensemble ${\mathcal{G}(N,L)}$ is defined by imposing the number of nodes $N$ 
and the number of edges $L$ as constraints on the ensemble and was introduced by the work of Paul Erdös and Alfréd Rényi \cite{erdds1959random}. 
The degree distribution of the graph follows the binomial distribution and its clustering coefficient tends to zero as the number of nodes increases,
whereas real-world networks tend to have a right-skewed degree distribution and higher clustering coefficient \cite{newman2018networks}. 
 
\subsubsection{Configuration model}
\label{microcanonical:cm}
The \gls{CM} is a widely recognized random graph model.  
It is a random graph model used to create simple graphs with a fixed degree sequence \cite{BOLLOBAS1980311}. 
In order to create a random graph with a fixed degree sequence, 
we begin by assigning a stub (also known as a half-edge) to each node for every one of its edges. 
Subsequently, we proceed to pair the stubs in a random manner, which results in a graph with a random structure. 
This yields a graph in which the degree sequence remains unchanged, while the rest of the graph is generated randomly. 
This methodology is commonly known as the stub-matching algorithm or the pairing model \cite{PhysRevE.64.026118,blitzstein2011sequential}. 
There are certain ramifications associated with this approach.  
One advantage of the \gls{CM} is its simplicity in implementation and ability to produce graphs with a fixed degree sequence. 
However, the \gls{CM} has a drawback in that it can produce graphs that contain self-loops and multiple edges, which may not always be desirable. 
One way to address the issue of self-loops and multiple edges is to simply eliminate them once the randomization procedure is completed. 
This is sometimes referred to as the \emph{erased configuration model} \cite{10.1214/12-SSY076} and 
can result in graphs that do not have the same degree sequence as the observed graph. 
An alternative strategy would involve eliminating the sampled graphs that contain self-loops and multiple edges, 
and retaining just the graphs that are simple. 
In the reference work of Newman \cite{newman2018networks}, 
it is suggested to allow self-loops and multiple edges when working with ``reasonably large networks'', 
since the likelihood of encountering self-loops and multiple edges decreases as the number of nodes increases.\\

The extension of the \gls{CM} to include directed graphs is described in \cite{10.1214/12-SSY076}. 
For directed graphs, we can employ the same methodology as for undirected graphs, 
but we must now differentiate between in-degrees and out-degrees. 
Given the observed graph's in-degree and out-degree sequence, 
denoted as $(\mathbf{k}_{\text{in}}, \mathbf{k}_{\text{out}})$, 
we can create a random graph by assigning each node $i$ with $k_{i,\text{out}}$ out-stubs and $k_{i,\text{in}}$ in-stubs. 
Subsequently, we proceed to randomly match the out-stubs with the in-stubs, which results in a random graph. 
Similar to the case of undirected graphs, this approach can result in graphs that contain self-loops and multiple edges. 
During the sampling process, there are two options available. 
The first option is to eliminate self-loops and multiple edges after the randomization process, which is known as the \emph{erased directed configuration model}. 
The second option is to discard the sampled graphs that contain self-loops and multiple edges.\\

An additional extension to partially directed graphs was introduced in \cite{partialdirecteconfmodel}. 
A partially directed graph consists of a combination of directed edges and undirected edges. 
In \cite{meyers2006predicting}, the authors applied a similar conceptual approach to directed networks, 
but this time they used probability generating functions to analyze what they called semi-directed networks. 
When joining the stubs, it is important to note that outgoing stubs can only be connected to incoming stubs, 
whereas undirected stubs can only be joined to other undirected stubs. 
To make sure the generated graph is simple, disconnected stubs, self loops, and parallel edges are removed.\\

Undirected bipartite graphs can be randomized using a similar approach to that used for directed graphs \cite{PhysRevE.64.026118}. 
From the observed degree sequences for each type of nodes ($\mathbf{k}_{U}$, $\mathbf{k}_{V}$), 
we may create a random bipartite graph by assigning a certain number of stubs to each node in $V$ and $U$. 
Matching the stubs between the different node groups then leads to the creation of a random bipartite graph. 
According to the definition of a bipartite graph, it is not possible to have self-loops, but we can encounter multiple edges. 
During the sampling procedure, we have two options: either we eliminate the multiple edges after the randomization phase, 
or we discard the sampled graphs that have multiple edges. 
The extension to directed bipartite graphs can be achieved in an analogous way.\\

From a practical point of view, the \gls{CM} is a very useful tool to generate random graphs with a specified degree sequence. 
As we have seen, the extensions to directed, bipartite and weighted graphs are not always straightforward and uniquely defined. 
Additionally, what is understood by \emph{``the configuration model''} might differ between references, 
which can cause confusion regarding the naming convention. 
The implementation of a software package's \emph{``the configuration model''} can also differ, which can cause more confusion.  
This observation was also made in \cite{differentconfigurationmodels}. 
It is important to understand the underlying random graph model, the implications of the assumptions made, 
and the algorithm used for generating the random graphs.

\subsection{Canonical ensembles}
\subsubsection{Random graph model ${\mathcal{G}(N,p)}$}
The random graph model ${\mathcal{G}(N,p)}$, defines the 
ensemble of graphs having $N$ nodes, where each edge is present with probability $p$ \cite{10.1214/aoms/1177706098}. 
It can be linked to the ${\mathcal{G}(N,L)}$ model by setting the expected number of links $\langle L \rangle$ as the ensemble constraint:
\begin{equation}
    \langle L \rangle = p \binom{N}{2} = \frac{N(N-1)}{2} p
\end{equation}
where $\binom{N}{2}$ is the number of possible edges in a graph of $N$ nodes. 
For a large value of $N$, the distinction between ${\mathcal{G}(N,p)}$ and ${\mathcal{G}(N,L)}$ 
becomes less important if $L$ is chosen to match the expected number of links in ${\mathcal{G}(N,p)}$.
The model suffers from the same drawbacks as the ${\mathcal{G}(N,L)}$ model.

\subsubsection{Chung-Lu model}
\label{ch:random:chunglu}
The Chung-Lu model is a random graph model that is used to generate graphs with a specified degree sequence \cite{chunglubase,chunglutwo}.
In the Chung-Lu model, the probability of an edge between two nodes $i$ and $j$ is given by \cite{osti_1239211}:
\begin{equation}
    p_{ij} = \begin{cases}
        \frac{k_i k_j}{2L} & \text{if } i \neq j; \\
        \frac{k_i^2}{4L} & \text{if } i = j
                \end{cases}
\end{equation}
and it is assumed that $ \max_i k_i^2 < \sum_i k_i$. 
This model also allows for self-loops. An in-depth analysis of the Chung-Lu model for undirected graphs can be found in \cite{osti_1239211}.
The Chung-Lu model can be extended to directed graphs as follows:
\begin{equation}
    p_{ij} = \begin{cases}
        \frac{k_{i,\text{out}} k_{j, in}}{L} & \text{if } i \neq j; \\
        \frac{k_{i,\text{out}}k_{i, in}}{2L} & \text{if } i = j.
                \end{cases}
\end{equation}
This model has been extended to bipartite graphs in \cite{DBLP:journals/corr/AksoyKP16}. 
In this case the probability of an edge between two nodes $i$ and $\mu$ is given by:
\begin{equation}
    p_{i\mu} = \frac{k_{i} k_{\mu}}{L}.
\end{equation}

\subsubsection{Maximum entropy models}
\label{ch:random:maxent}
In applications, it is advantageous to not only have an ensemble to represent the system being studied, 
but also to have a graph ensemble that is unbiased and uses the available information without making any additional assumptions a priori \cite{ginestrabianconi2018}. 
The maximum-entropy method involves constructing an ensemble of random graphs, denoted as $\mathcal{G}$, based on a given observed graph $G^{*}$. 
The graphs in $\mathcal{G}$ have a random topology, except for a specific set of structural constraints, 
denoted as $\mathbf{C}$, which are measured on $G^*$ \cite{Squartini_2011,Squartini_2015,squartini2017maximum-entropy,Garlaschelli_2008}. 
The least-biased  ensemble is found by maximizing the Shannon entropy $S$ :
\begin{equation}
    S = - \sum_{G\in \mathcal{G}} P(G) \ln{ P(G)},
\end{equation}
subject to the set of constraints $\mathbf{C}$, which are imposed on average over the ensemble 
and subject to the normalization condition (Equation \ref{eq:random:sumprob}).
This constrained optimization problem can be solved by maximizing the associated Langrangian function 
with respect to $P(G)$ \cite{PhysRevE.70.066117}.
From the maximum entropy probability distribution associated with some set of constraints, 
we can obtain the maximum entropy ensemble corresponding with the observed graph $G^{*}$. 

\paragraph{Different models}
The \gls{UBCM} is the canonical equivalent of the microcanonical configuration model from Section \ref{microcanonical:cm}. 
In this model we impose the degree sequence $\mathbf{k}^*$ of the observed graph $G^*$ as a constraint \cite{PhysRevE.70.066117,Squartini_2011}.
The \gls{DBCM} is the extension of the \gls{UBCM} to directed graphs.
In this model, both the in- and the outdegree sequence of the observed graph $G^*$ are imposed as constraints \cite{PhysRevE.70.066117,Squartini_2011}.
In the case of the \gls{BiCM}, the degree sequence of both sets of nodes of the observed bipartite graph $G^*$, 
$\mathbf{k}_U$ and $\mathbf{k}_V$ are imposed as constraints \cite{Saracco:2015aa}.
The previously cited models are merely a subset of the maximum entropy models that exist in the literature. 
Even when focusing solely on simple networks, there are still numerous other models that may be taken into account. 
For additional examples, we refer to \cite{PhysRevE.73.015101,Squartini_2011,2014,2016,bianconi2013,realstats}

\section{Algorithmic approaches to graph randomization}
The impact of the underlying algorithm for generating random graphs with prescribed degree sequences has been studied over 20 years ago \cite{milouniformgraphs}. 
The authors conducted a comparison of three distinct algorithms (stub-matching, edge switching, and a Monte Carlo procedure) 
in order to demonstrate the impact of the algorithm and its ability to sample the graph ensemble in an unbiased manner. 
This section will look at some more details about the variety of algorithms that can be used to create random graphs. 
This section does not provide a comprehensive overview of all available algorithms. 
Instead, it seeks to provide an overview of the algorithms used in popular software packages described later. \\

The presence or absence of self-loops and multiple edges are often overlooked. 
Nevertheless, it is important to realize that the existence of self-loops and many edges might greatly affect the outcomes of the study. 
In \cite{differentconfigurationmodels}, the concept of a random graph with a fixed degree sequence is shown to be applicable to eight overlapping, 
yet distinct graph spaces. 
The most suitable graph space can be determined based on the presence or absence of self-loops, multiple edges, and stub labels in the graph being studied. 
Users should also note that the existence of self-loops and many edges can have an effect on the outcomes of downstream tasks. 
For example, does the procedure for counting subgraphs consider the existence of self-loops and multiple edges? 
If it is the case, in what manner can it be done? Is this behavior considered desirable, and to what extent is it documented in the tools being used?

\subsection{Chung-Lu}
\label{ch:random:chunglu:algo}
In order to create a random graph with a given degree sequence using the Chung-Lu model, the simplest method is to directly apply the Chung-Lu model. 
This involves sampling a random number between 0 and 1 for each pair of nodes $i$ and $j$, and comparing it to the edge probability $p_{ij}$. 
This results in an algorithm with a time complexity of $\mathcal{O}(N^2)$. 
The same principle can be used for directed and bipartite graphs as well. 
A more efficient method that operates with a time complexity of $\mathcal{O}(L k_{\text{max}})$, 
where $k_{\text{max}}$ represents the maximum degree of the graph has been suggested in \cite{networkxchunglu}. 
The graph is constructed by incrementally adding edges, with the nodes selected in proportion to the degree distribution. 
The degree distribution is updated iteratively after each edge is added. 
Specifically, when a node is selected, its degree is reduced by one for the subsequent iteration after the edge is added. 
The chosen nodes must be distinct from one another, and it is prohibited to add duplicate edges. 
A different approach that runs in $\mathcal{O}(N + L)$ is proposed in \cite{10.1007/978-3-642-21286-4_10}. 
Starting from the nodes ranked by degree, the algorithm adds edges between the nodes that are likely to be connected and 
has the ability to skip over nodes that have a low probability of being connected.
By jumping over unlikely connections, it can build the network much faster than if it had to consider every possible connection.
In \cite{DBLP:journals/corr/abs-2103-00662} the ``shifted Chung-Lu'' method is proposed, 
which aims to control the variations in the degree distributions of randomization graphs. 
This is achieved by employing a Poisson approximation to establish a linear system that describes the predicted degree sequence. 
The authors employed Matlab for their experiments, and as far as we are aware, this method has not yet been included into any of the tools mentioned.
An approximate algorithm for Chung-Lu graphs also exists for bipartite graphs \cite{10.5555/982792.982902}.

\subsection{Stub-matching}
\label{ch:random:stubs}
\begin{figure}[p]
  \centering
  \includegraphics[width=0.75\textwidth]{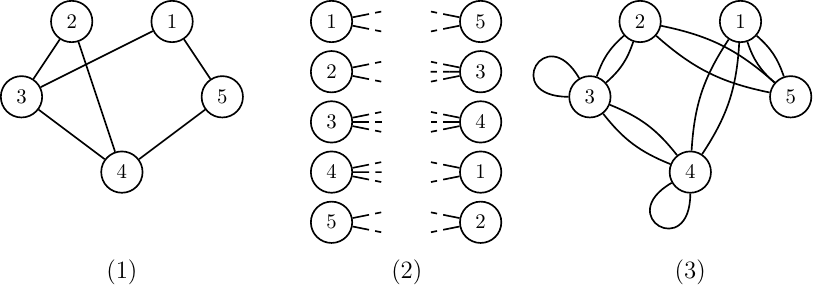}
  \caption[Stub-matching algorithm]{Randomization using stub-matching on a simple graph. (1) original graph (2) half graphs to be combined with another half graph (3) resulting graph (including self-loops and multiple edges).}
  \label{fig:randomization:stubs}
  \Description{Randomization using stub-matching on a simple graph.}
\end{figure}

The principles behind the stub-matching method were explained in the preceding section on the configuration model. 
Overall, it has the benefit of being straightforward to implement and efficient in terms of speed. 
As previously stated, it has the potential to produce graphs that contain self-loops and multiple edges, which may not always be desirable. 
By eliminating self-loops and multiple edges during the randomization process, 
the resulting graph may not have the identical degree sequence as the original graph. 
Consequently, we are no longer uniformly sampling from the ensemble of graphs with the same degree (or strength) sequence(s). 
Fortunately, as the number of nodes increases, the likelihood of self-loops and multiple edges decreases considerably \cite{newman2018networks}. 
By excluding graphs that contain self-loops and multiple edges, we can effectively sample from a collection of graphs that share the same degree sequence(s). 
However, this process may be time-consuming in the case of heterogeneous degree distributions \cite{milouniformgraphs}. 
Figure \ref{fig:randomization:stubs} provides a visual representation of the stub-matching algorithm.\\

\subsection{Switching algorithms}
\label{ch:random:switching}
\begin{figure}[p]
    \centering
    \includegraphics[width=0.5\textwidth]{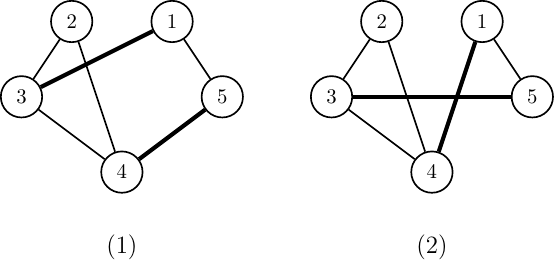}
    \caption[Link rewiring algorithm]{Randomization using \gls{LRA} on a simple graph with the switching edges in bold (1) original graph (2) \gls{LRA} single rewiring step, maintaining the degree sequence.}
    \label{randomization:LRA}
    \Description{Randomization using \gls{LRA} on a simple graph.}
\end{figure}

The switching algorithm, also referred to as the \gls{LRA} \cite{maslov2002specificity}, 
can be used to create a random graph with a prescribed degree sequence. 
The idea of the algorithm is to randomly select two edges and to swap their endpoints, 
whilst assuring that the swap does not create self-loops or multiple edges. 
When done in this way, the algorithm is able to sample  the ensemble uniformly, provided that the number of switches is large enough.
In \cite{milouniformgraphs}, using a total of $QL$ switches is suggested, 
where $L$ represents the number of edges and $Q$ should be sufficiently large (a proposed value is $Q=100$). 
Figure \ref{randomization:LRA} provides a visual representation of the \gls{LRA}.\\

In addition to the \gls{LRA} method, the authors of \cite{Gkantsidis2003TheMC} introduced a Markov Chain simulation approach 
for creating connected graphs with a specified degree sequence. 
In order to enhance the efficiency of the method, a heuristic is used to minimize the frequency of connectivity checks for the graph following a swap.  
Swaps are made within a window of $T$ steps without checking for connectivity. 
If the graph obtained after performing these steps is connected, the size of the window is expanded; 
otherwise, it is decreased. 
By performing this, it is expected that the variable $T$ would spontaneously adapt in order to strike a balance between 
a sizable window $T$ and a high likelihood of the graph being connected after $T$ swaps. 
The window size $T$ was optimized in \cite{DBLP:journals/compnet/VigerL16} to further enhance this strategy. 
This enhancement enables the generation of connected graphs with a specified degree sequence at a faster rate.
Ongoing research continues to explore more algorithms that are currently not included in various software packages. 
For example, in \cite{mannion2024fast}, instead of rewiring only two edges at a time, 
the authors suggest rewiring several edges simultaneously and allow for tuning of the network's assortativity.

\subsection{Curveball}
\label{ch:random:curveball}
\begin{figure}[p]
    \centering
    \includegraphics[width=0.75\textwidth]{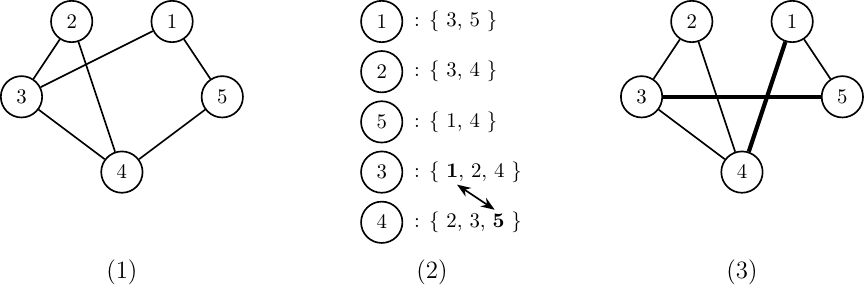}
    \caption[Curveball algorithm]{Randomization using the curveball algorithm on a simple graph (1) original graph (2) neighbor lists and single round of swaps represented by the arrows (3) resulting graph.}
    \label{randomization:curveball}
    \Description{Randomization using the curveball algorithm on a simple graph.}
\end{figure}

The primary concern associated with switching algorithms is that they do not guarantee reliable outcomes for large graphs 
unless the number of swaps is increased to a level that renders the computation itself quite problematic. 
In order to address this issue, the \emph{curveball method} was introduced \cite{curveballmethod}. 
The initial design of the algorithm was meant to work with biadjacency matrices, 
however it is also capable of working with adjacency matrices representing simple (un)directed graphs \cite{CARSTENS2018773}. 
The concept behind the algorithm is as follows: Initially, we create the neighbor list for every node. 
Next, we proceed to randomly choose nodes and swap their disjoint neighbors (referred to as ``trading''). 
The number of transactions between the two nodes ranges from 0 to the smaller of the two list lengths. 
When comparing the switching algorithm with the curveball approach, the latter is generally faster. 
Figure \ref{randomization:curveball} shows an example of the curveball approach. 
In this example, the neighbor lists of node $3$ and $4$ are chosen and they swap nodes $1$ and $5$ respectively.

\subsection{Sampling from a maximum entropy ensemble}
\label{Ch3:random:sampling}
For the various maximum entropy ensembles mentioned in section \ref{ch:random:maxent}, and in the case of unweighted graphs, 
the elements of the adjacency matrix are independent and have only two possible values.  
Therefore we may use the predicted value of the adjacency matrix to sample from the ensemble. 
Each entry in the adjacency matrix is a Bernoulli random variable, 
with its parameters determined by the Lagrange multipliers that maximize the likelihood. 
By iterating over all unique elements of the (bi)adjacency matrix, we can generate a random graph from the ensemble.

\section{Statistical significance}
\label{Ch:random:significance}
In the preceding section, we have examined various approaches for graph randomization and have studied specific models in more detail. 
Being able to create random graphs, we may now use them for assessing the statistical significance of a metric or a topological feature of a network. 
Hypothesis testing, in the context of classical statistics, 
is a method that involves making decisions about population parameters using data obtained from a population sample. 
The $p$-value is an essential part of this procedure, functioning as a metric to quantify the level of evidence against the null hypothesis. 
The $p$-value is the probability of obtaining a result that is as extreme or more extreme than the observed result, 
under the assumption that the null hypothesis is true \cite{Ross_2021}.\\

In practice, the following steps are typically taken: a research question is formulated, 
and a null hypothesis $H_0$ and an alternative hypothesis $H_1$ are defined. 
For instance, 
we may like to determine whether the assortativity coefficient of our observed network is lower than the expected value for a given null model. 
Next, we calculate the assortativity coefficient for both the observed network and a sample of random networks. 
We then we compare the observed assortativity coefficient to the distribution of the metric in the sample. 
Based on this analysis, we draw a conclusion.  
The main concept behind this approach is to compare the observed value of a metric with the metric's distribution in the ensemble of random graphs.

\subsection{Quantifying the significance of a metric}

Generating many samples from the ensemble can be seen as a Monte Carlo procedure and 
may be used to determine an empirical $p$-value that approximates the actual $p$-value without depending on or having knowledge of the metric's distribution. 
The empirical $p$-value is typically defined as the proportion of randomized graphs that exhibit a metric value that is equal to 
or more extreme than the observed value:

\begin{equation}
    \frac{1}{N_{rand}} \sum_{i=1}^{N_{rand}} \mathbb{I}(N_{real} \geq N_{i}), 
\end{equation}
where $N_i$ denotes the value of the metric in random graph $i$ and 
$\mathbb{I}$ is the indicator function that returns 1 if the condition is true and 0 otherwise.
Following the recommendations in \cite{exppvale}, we will use the following definition of the empirical $p$-value, $\hat{p}$ (in the case of a right-tailed test):
\begin{equation}
    \hat{p} = \frac{1}{N_{rand} + 1} \left( 1 + \sum_{i=1}^{N_{rand}} \mathbb{I}(N_{real} \geq N_{i}) \right).
\end{equation}
To quantify the importance or significance, the $z$-score is often used as well:
\begin{equation}
    z = \frac{X_{real} - \langle X_{\text{rand}} \rangle}{\sigma_{\text{rand}}},
\end{equation}
where $\sigma_{rand}$ denotes the standard deviation of the metric $X$ in the randomized graphs. 
The $z$-score is the divergence of the metric in the observed graph from the average of the randomized graphs, 
measured in terms of standard deviations. 
If the $z$-score exceeds (or falls below) a specific threshold, the metric is considered to be statistically significant\footnote{Depending on considering a left-tailed, right-tailed or two-tailed hypothesis test.}. 
An inherent drawback of this particular interpretation of the $z$-score is that it assumes the metric adheres to a normal distribution. 
When dealing with variables that do not follow a normal distribution, it is possible for large values to still indicate the most deviating patterns \cite{squartini2017maximum-entropy}. 
However, it is important to exercise caution, as we will discuss further. 
The literature occasionally mentions this aspect of the assumed normalcy of the metric, although it is often disregarded or overlooked \cite{Neal:2023gol}.

\subsection{Typical applications}
\subsubsection{Network motifs}
\begin{figure}
  \centering
  \includegraphics[width=0.75\textwidth]{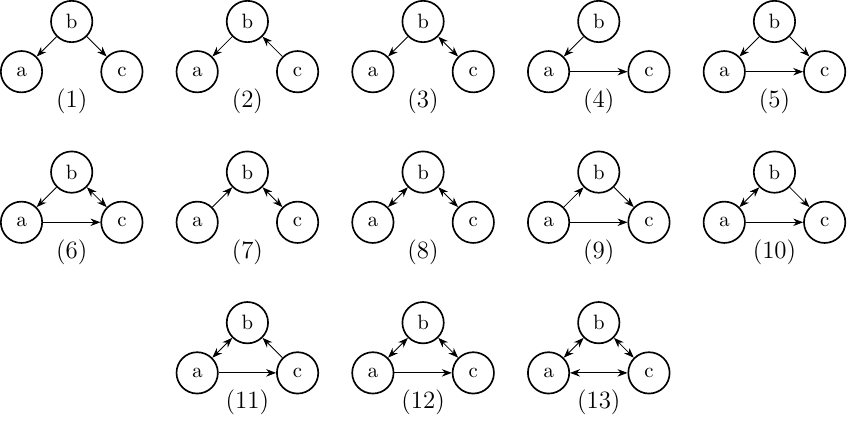}
  \caption[Unique three-node directed subgraphs]{The 13 unique patterns of directed subgraphs involving three nodes for unweighted, directed graphs.}
  \label{fig:basics:directedsubgraphs}
  \Description{}
\end{figure}

\begin{figure}
  \centering
  \includegraphics[width=0.7\textwidth]{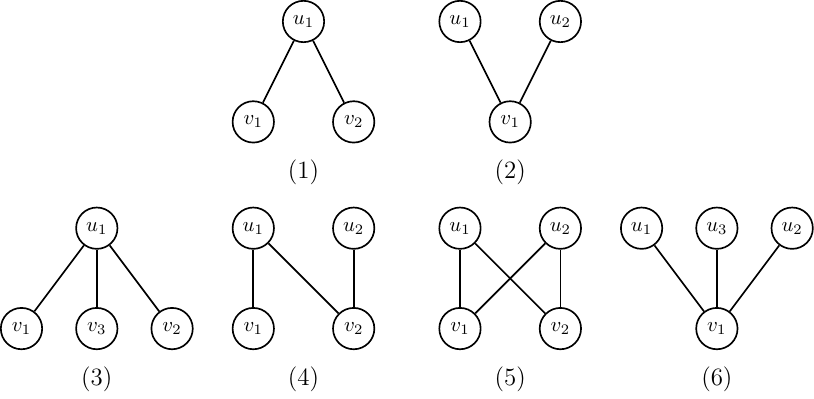}
  \caption{Common subgraphs for undirected bipartite graphs.}
  \label{fig:basics:bipartitepatterns}
  \Description{}
\end{figure}

The notion of ``network motifs'' to analyze the statistical importance of subgraph occurrences in graphs was introduced in \cite{motifbasicsmilo2002}. 
Network motifs refer to patterns of interconnections that appear in graphs at frequencies much higher than those seen in randomized versions of the same graphs. 
Figure \ref{fig:basics:directedsubgraphs} shows common subgraphs for directed graphs involving three nodes 
and Figure \ref{fig:basics:bipartitepatterns} shows common subgraphs for undirected bipartite graphs involving three or four nodes.

The original publication on network motifs established three requirements for a subgraph to be classified as a motif. 
The probability $p$ of the subgraph occurring in the randomized graphs must be less than a specified threshold, often $p < 0.05$. 
Furthermore, the subgraph must occur at least four times in the observed graph. 
Let $N_{real}$ denote the number of subgraphs in the observed graph, and $\langle N_{rand} \rangle$ represent the average number of subgraphs in the randomized graphs. 
The third criterion requires that the number of occurrences in the observed network should be substantially greater than the number of occurrences in the randomized networks: 
$N_{real} - \langle N_{rand} \rangle > 0.1 \langle N_{rand} \rangle$. 
This is to prevent labelling a subgraph with only a small difference, but a narrow distribution in the ensemble as a motif. 
The $z$-score of a motif is defined as \cite{vitolatora2017}:
\begin{equation}
    z = \frac{N_{real} - \langle N_{rand} \rangle}{\sigma_{rand}},
\end{equation}
where $\sigma_{rand}$ denotes the standard deviation of the number of subgraphs in the randomized graphs.\\

The empirical $p$-value and the z-score are two statistical indicators measures associated with motifs. 
A comprehensive summary of supplementary statistical measures that can be used to quantify the significance of network motifs is provided in \cite{motifsurvey}. 
The authors additionally differentiate between statistical measures and structural measures. 
The latter category encompasses metrics that combine motifs with graph metrics. 
For the rest of this paper, our main focus will be on statistical measures. \\

When performing many hypothesis tests, the likelihood of making at least one false discovery increases substantially. 
Consequently, the FDR process can employed to determine the statistical significance of each $p$-value in such cases \cite{10.2307/2346101}. 
For some applications such as the projection of bipartite graphs, 
using the FDR process on the $p$-values of all edges in the projected network can be computationally expensive. 
In these situations, a linear time variant of the FDR process can be employed as an alternative \cite{10.1093/bioinformatics/btw029}.

\subsubsection{Validating the projection of a bipartite graph}
\label{ch:random:bipartiteprojection}
A bipartite graph $G$ can be represented by its $N_V \times N_U$ biadjacency matrix $\mathbf{B}$ with elements $b_{i\mu}$, 
where $N_V$ and $N_U$ represent the number of nodes in the two sets of nodes $V$ and $U$ respectively. 
A monopartite graph $G'$ can be obtained from $G$ by projecting the bipartite graph onto one of the node sets (or layers).
In the case where we want to count the number of times two nodes are connected in the projection, the resulting monopartite graph $G'$ will be weighted. 
For a projection on the node set $U$ the elements of the adjacency matrix $\mathbf{A}'$ of $G'$ are defined as follows:
\begin{equation}
    a'_{\mu \nu} = \begin{cases}
                0 & \text{if $\mu = \nu$};\\
                \sum_{i \in V} b_{i\mu} b_{i\nu} & \text{otherwise}.\\    
\end{cases}
\end{equation}
Due to the shape of these patterns, they are often referred to as $\mathcal{V}$-motifs (cf. Figure \ref{fig:basics:bipartitepatterns} (1) and (2)).\\

\begin{figure}
  \centering
  \includegraphics[width=0.7\textwidth]{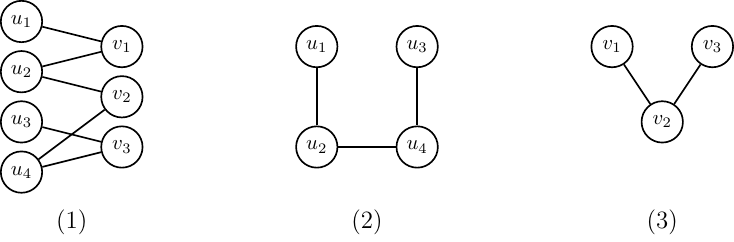}
  \caption[Example of a bipartite graph and its projection onto its node sets.]{Example of a bipartite graph and its projection onto its node sets. (1) Original bipartite graph (2) projection on node set $U$ (3) projection on node set $V$}
  \label{basics:bipartiteprojection}
  \Description{Example of a bipartite graph and its projection onto its node sets.}
\end{figure}

In the case where we simply want to know whether two nodes are connected in the projection, 
the resulting monopartite graph $G'$ is unweighted.
For a projection on the node set $U$ the elements of its adjacency matrix $\mathbf{A}'$ are defined as follows:
\begin{equation}
    a'_{\mu \nu} = \begin{cases}
                0 & \text{if $\mu = \nu$};\\
                \theta \left( \sum_{i \in V} b_{i\mu} b_{i\nu} \right) & \text{otherwise},\\    
\end{cases}
\end{equation}
where $\theta$ is the Heaviside step function. 
Figure \ref{basics:bipartiteprojection} shows an example of an undirected bipartite graph and its projection onto its node sets.\\

We can assess the statistical significance of each edge in the projected network by calculating the $p$-value of the edge. 
In particular, we compare the observed number of $\mathcal{V}$-motifs with the distribution of the number of $\mathcal{V}$-motifs in the ensemble of random graphs. 
By performing this procedure for every edge in the projected network, 
we can obtain the $p$-value for each edge in the projected network. 
This is the approach followed in \cite{bipartite-meta} and this validation procedure is commonly known as backbone extraction. 
A comprehensive analysis of the validation of the projection of a bipartite graph using different null models is presented in \cite{Neal:2021aa}. 
The authors suggest using the \gls{BiCM} to derive the probabilities of the biadjacency matrix of the randomized graph for its superior speed and accuracy.
If an acceptable significance level is selected, the microcanonical ensemble can also identify similar backbones. 
An approach to reconcile different null models for the projection of a bipartite graph is presented in \cite{bipartite-meta}. 
Here the authors focus on setting the significance level for each null model in such a way that the density of validated edges is similar. 
The choice of network density is motivated by the idea that statistical validation of networks typically aims at highlighting the emergence of patterns and communities.\\

In the context of the \gls{BiCM}, the number expected $\mathcal{V}$-motifs follows a Poisson-binomial distribution \cite{Becatti:2019aa}. 
Using the maximum entropy approach, we can compute the expected number of $\mathcal{V}$-motifs between a node $i$ and $j$ from the set of nodes $V$ as follows:
\begin{equation}
    \langle a'_{ij} \rangle = \sum_{\mu \in U}  p_{i\mu}p_{j\mu},
\end{equation}
where $p_{i\mu}$ denotes the probability of an edge between nodes $i$ and $\mu$ under the \gls{BiCM}, and $a'_{ij}$ denotes the observed number of $\mathcal{V}$-motifs between nodes $i$ and $j$. 
The elements of the sum can be considered to be the parameters of a Poisson-binomial distribution. We can calculate the $p$-value of the edge as follows:
\begin{equation}
    p_{a'_{ij}} = \sum_{a'_{ij} \ge {a'_{ij}}^{*}} f_{PB}(a'_{ij}),
\end{equation}
where $f_{PB}(a'_{ij})$ denotes the probability mass function of the Poisson-binomial distribution. 
For practical application where $\langle a'_{ij} \rangle$ is small, the Poisson approximation of the Poisson-binomial distribution can be used. 
For specific choices of what the layers actually represent, 
it is even possible to determine the parameters without having to explicitly solve the system of equations \cite{Becatti:2019aa}.\\ 

\section{The tooling landscape}
We have observed that there are many methods for randomizing graphs, each with its own unique advantages and disadvantages. 
Gaining a proper understanding of the various methodologies and fundamental algorithms is a first step.  
However, in practical situations, researchers often rely on existing tools to perform the randomization process and seldom implement an algorithm from scratch. 
This section will provide an overview of commonly used tools for graph randomization and their functionalities. 
In the following section, we will then use these tools to generate random networks and evaluate and compare the results. 
Table \ref{chrand:tab:tools} presents an overview of the tools used, along with their respective versions. 
In what follows, we will present a more in-depth analysis of the different tools, 
with a specific emphasis on randomization methods that involve the fixing of degree sequences. 
Given the nature of the entropy-based approaches, we distinguish them from the rest of the tools. \\

\begin{table*}
  \centering
  \caption[Available tools for graph randomization]{Tools for graph randomization.}
  \label{chrand:tab:tools}
  {\footnotesize
    \begin{tabular}{lll} 
      \toprule
      Tool & General description & Version\\ 
      \midrule
      \parbox[t]{2.5cm}{\emph{NetworkX} \cite{SciPyProceedings_11}} & \parbox[t]{8cm}{A Python module for the creation, manipulation, and study of the structure, dynamics, and functions of complex networks.} &  \parbox[t]{2cm}{v3.2.1 with \\Python 3.11.5}\\
      \parbox[t]{2.5cm}{\emph{igraph} \cite{igraph,igraphmanual}}   & \parbox[t]{8cm}{A collection of network analysis tools with emphasis on efficiency, portability and ease of use (available for R, Python, Mathematica and C/C++).} & \parbox[t]{2cm}{v0.11.3 with \\Python 3.11.5}\\
      \parbox[t]{2.5cm}{\emph{Networkit} \cite{DBLP:journals/corr/StaudtSM14}} & \parbox[t]{8cm}{A Python module for open-source toolkit for large-scale network analysis (with the performance-aware algorithms written in C++).} &  \parbox[t]{2cm}{v10.1 with \\Python 3.11.5}\\
      \parbox[t]{2.5cm}{\emph{graph-tool} \cite{peixoto_graph-tool_2014}} & \parbox[t]{8cm}{A Python module for manipulation and statistical analysis of graphs (the core data structures and algorithms are implemented in C++).} &  \parbox[t]{2cm}{v2.58 with \\Python 3.11.5}\\
      \parbox[t]{2.5cm}{\emph{NEMtropy} \cite{Vallarano:2021ws}} & \parbox[t]{8cm}{A Python module for generating random networks based on the maximum entropy principle.} &  \parbox[t]{2cm}{v2.1.1 with \\Python 3.11.5}\\
      \parbox[t]{2.5cm}{\emph{MaxEntropyGraphs.jl} \cite{BDC_MEG_latest}} & \parbox[t]{8cm}{A Julia package for generating random networks based on the maximum entropy principle.} &  \parbox[t]{2cm}{v0.4.3 with Julia 1.9}\\
      \bottomrule
    \end{tabular}
  }
\end{table*}

The software ``bmotif'' enables the analysis of motifs in bipartite graphs \cite{https://doi.org/10.1111/2041-210X.13149}. 
While Table \ref{chrand:tab:tools} does not include it due to its lack of support for randomization, 
this tool can still be used to examine the results of randomization techniques. 
It was especially built for the R programming language, but it may also be used with Matlab and Python. 
The software enables the enumeration of specific subgraphs in bipartite graphs. 
Since this software was specifically designed for ecological networks, 
it also enables the calculation of species occurrence frequency inside motifs.\\

Table \ref{tbl:chrand:comparison} summarizes the functionality of each tool, categorized by conceptual approach and algorithm. 
From the table, it is clear that the number of randomization methods and algorithms significantly decreases as graph complexity increases. 
For example, while there is a version of the Chung-Lu model that can be applied to directed graphs, 
none of the tools have incorporated this feature. 
Similarly, none of the tools incorporate the curveball algorithm for bipartite graphs, 
although the method was originally designed for such graphs. 
This limitation gets even more severe when dealing with weighted graphs, 
requiring researchers to implement the algorithm(s) to meet their particular needs. 
The following paragraphs offer additional details about each tool.

\begin{table*}
  \centering
  \caption[Package comparison by algorithm availability]{Comparison of the functionality and the available algorithms for the standard tools considered in this study. The references for the underlying algorithms are indicated wherever applicable.}
  \label{tbl:chrand:comparison}
  {\footnotesize
  \begin{tabular}{p{1.25cm}p{2.65cm}p{2.65cm}p{3cm}p{3cm}}
      \toprule
      & \textbf{Stub-matching} & \textbf{LRA} & \textbf{Chung-Lu} & \textbf{Curveball} \\
      \midrule
      \multicolumn{5}{l}{\textbf{Undirected}} \\
      NetworkX  & \parbox[t]{2.65cm}{\texttt{configuration\_model}\textsuperscript{1,2}} 
                & \parbox[t]{2.65cm}{\texttt{random\_reference}\textsuperscript{3} \cite{maslov2002specificity}}
                & \parbox[t]{3cm}{\texttt{expected\_degree\_graph}\textsuperscript{2} \cite{10.1007/978-3-642-21286-4_10}}
                & \parbox[t]{3cm}{\texttt{random\_degree\_ sequence\_graph}\textsuperscript{4,5} \cite{networkxchunglu}}\\
      iGraph    & \parbox[t]{2.65cm}{\texttt{Degree\_Sequence}\\ (configuration)\textsuperscript{1,2}\\
                                  \texttt{Degree\_Sequence}\\ (configuration\_simple)\\
                                  \texttt{Degree\_Sequence}\\ (fast\_heur\_simple)\textsuperscript{5}}
                & \parbox[t]{3cm}{\texttt{Degree\_Sequence}\\ (edge\_switching\_simple)\textsuperscript{4*}\\
                                  \texttt{Degree\_Sequence}\\ (vl) \cite{DBLP:journals/compnet/VigerL16}}
                & None 
                & None\\
      Networkit & None
                & \parbox[t]{2.65cm}{\texttt{Configuration ModelGenerator} \cite{DBLP:journals/compnet/VigerL16}\\
                                  \texttt{EdgeSwitching} \cite{Gkantsidis2003TheMC}}
                & \parbox[t]{2.65cm}{\texttt{ChungLuGenerator}\textsuperscript{**} \cite{10.1007/978-3-642-21286-4_10}}
                & \parbox[t]{3cm}{\texttt{Curveball} \cite{curvbalnetworkit}\\
                                  \texttt{GlobalCurveball} \cite{curvbalnetworkit}}\\
      graph-tool& None
                & \parbox[t]{2.65cm}{\texttt{random\_graph}\textsuperscript{1,2}}
                & None
                & None\\

      \midrule
      \multicolumn{5}{l}{\textbf{Directed}} \\
      NetworkX  & \parbox[t]{2.65cm}{\texttt{directed\_ configuration\_model}\textsuperscript{1,2}} 
                & None
                & None
                & None\\
      iGraph    & \parbox[t]{2.65cm}{\texttt{Degree\_Sequence}\\ (configuration)\textsuperscript{1,2}\\
                                  \texttt{Degree\_Sequence}\\ (configuration\_simple)\\
                                  \texttt{Degree\_Sequence}\\ (fast\_heur\_simple)\textsuperscript{5}}
                & \parbox[t]{3cm}{\texttt{Degree\_Sequence}\\ (edge\_switching\_simple)\textsuperscript{4}}
                & None 
                & None\\
      Networkit & None
                & \parbox[t]{2.65cm}{\texttt{EdgeSwitching} \cite{Gkantsidis2003TheMC}}
                & \parbox[t]{3cm}{None}
                & \parbox[t]{3cm}{\texttt{GlobalCurveball} \cite{curvbalnetworkit}}\\          
      graph-tool& None
                & \parbox[t]{3cm}{\texttt{random\_graph}\textsuperscript{1,2}}
                & None
                & None\\
      \midrule
      \multicolumn{5}{l}{\textbf{Bipartite}} \\
      NetworkX  & \parbox[t]{2.65cm}{\texttt{configuration\_model}\textsuperscript{1,2}} 
                & None
                & None
                & None\\
      \bottomrule
      \multicolumn{5}{l}{\parbox[t]{13cm}{\textsuperscript{1} Multiple edges can occur\\
                                          \textsuperscript{2} Self-loops can occur\\
                                          \textsuperscript{3} The graph can be connected or not\\
                                          \textsuperscript{4} The algorithm can fail to produce a random graph\\
                                          \textsuperscript{5} The sampling from the ensemble is non-uniform.\\
                                          \textsuperscript{*} This function will call upon the \texttt{rewire} method, which by default only does $1000$ rewiring trials, whereas according to \cite{milouniformgraphs}, around 100 times the number of edges is recommended for proper mixing.\\
                                          \textsuperscript{**} The nodes of the returned random graph are sorted according to their degree in the original graph.}} \\
  \end{tabular}
  }
\end{table*}

\subsection{NetworkX}
NetworkX is a Python module that allows for the creation, modification, 
and analysis of complex networks in terms of their structure, dynamics, and functions \cite{SciPyProceedings_11}. 
The software is compatible with directed, undirected, and bipartite graphs, as well as multigraphs (graphs with parallel edges). 
NetworkX offers many options for generating random graphs based on a specified degree sequence. 
The stub-matching, \gls{LRA}, and Chung-Lu methods are all available. 
If self-loops or multiple edges are present, the user can choose to eliminate them, 
however this may result in a graph that does not match the prescribed degree sequence.

\subsection{Igraph}
Igraph is a suite of network analysis tools that prioritizes efficiency, portability, and user-friendliness. 
It is available for use with R, Python, Mathematica, and C/C++ \cite{igraph,igraphmanual}. 
Igraph offers a variety of functions to create random graphs with a specified degree sequence. 
These functions can generate both directed and undirected graphs, and they encompass different approaches.
Like NetworkX, many functions can generate graphs with multiple edges and self-loops, 
which can result in a graph that does not precisely match the specified degree sequence. 
There is no built-in capability to use the Chung-Lu model or to handle bipartite graphs.

\subsection{Networkit}
Networkit is a Python toolkit for analyzing large-scale networks. 
It is open-source and includes performance-aware algorithms written in C++ \cite{DBLP:journals/corr/StaudtSM14}. 
Although it lacks the stub-matching approach, this tool does provide the \gls{LRA} and Chung-Lu approaches. 
Additionally, it is the sole tool that incorporates the curveball method described in Section \ref{ch:random:curveball}. 
When using a Chung-Lu model, it is crucial to note that the nodes of the resulting random graph are arranged based on their degree in the initial network. 
This is particularly important to be aware of when examining metrics at the node level. 
Both directed and undirected graphs are supported, but the tool does not have any built-in capabilities to handle bipartite graphs.

\subsection{Graph-tool}
Graph-tool is a Python module that allows for the manipulation and statistical analysis of graphs. 
Its main data structures and algorithms are implemented in C++ \cite{peixoto_graph-tool_2014}. 
The tool offers features for generating random graphs with a specified degree sequence. 
It uses a \gls{LRA} approach described in Section \ref{ch:random:switching}. 
The user has the option to include or exclude self-loops and multiple edges. 
There is no available feature to create bipartite graphs with a predetermined degree sequence. 
Graph-tool offers a wide range of additional models, including blockmodel ensembles \cite{PhysRevE.83.016107,HOLLAND1983109}, 
as well as ensembles that preserve in- and out-degree correlations. 
To keep things concise, we shall refrain from discussing or using these models in this context. 

\subsection{NEMtropy}
NEMtropy is a Python module that uses the maximum entropy principle to generate random graphs \cite{Vallarano:2021ws}. 
It can be used to compute the likelihood maximizing parameters for the \gls{UBCM}, the \gls{DBCM}, and the \gls{BiCM} (among other models). 
The package includes an ensemble sampler that can be used to generate random graphs. 
The samplers have been implemented in accordance with the explanation provided in Section \ref{Ch3:random:sampling} 
and are thoroughly discussed in \cite{Vallarano:2021ws}. 
The graphs produced by the samplers do not contain self-loops or multiple edges. 
One drawback of the NEMtropy samplers is that they store the created graphs on disk as edgelists, which require additional processing. 
In contrast, most other tools generate the graphs directly in memory.

\subsection{MaxEntropyGraphs.jl}
MaxEntropyGraphs.jl is a Julia package that creates random networks using the maximum entropy principle \cite{BDC_MEG_latest}. 
It bears resemblance to NEMtropy, but is coded in Julia.
It has the capability to compute the likelihood maximizing parameters for the \gls{UBCM}, the \gls{DBCM}, the \gls{BiCM}, and other models.  
The package includes an ensemble sampler that can be used to generate random graphs. 
The software also offers supplementary features for analyzing the generated graphs. 
For example, users can seamlessly integrate their own network metrics. 
Functionalities to compute count the occurrence of subgraphs and to extract the backbone of a bipartite graph are also available.
The samplers are implemented following the methods described in Section \ref{Ch3:random:sampling}. 
The graphs produced by the samplers do not contain self-loops or multiple edges.

\section{Comparative Analysis and Discussion}
\label{sec:comparison}
Currently, we possess a solid understanding of various graph randomization algorithms, 
the approaches for assessing the statistical significance of a metric or topological feature of a network, 
and the tools that allow us to perform the randomization process. 
In this section, we consolidate every aspect: we use the tools to create random graphs, and subsequently analyze and compare the outcomes. 
It can be argued that this section partially addresses two separate topics. 
On one hand, we are comparing the results obtained from different randomization methods, while on the other hand, 
we are comparing the conclusions derived from distinct families of null models, since we are evaluating models that have both hard and soft constraints. 
The concept of comparing various null models is not novel per se, 
but researchers should be aware of the influence that the underlying algorithm can have 
on the conclusions derived from the randomization process.\\

Several studies have examined the outcomes derived from different randomization techniques in the context of bipartite graphs.
An in-depth review of null models for bipartite networks can be found in \cite{Neal:2023gol}. 
The matter of selecting a suitable null model is also discussed. 
From an algorithmic standpoint, it is recommended to select the option with the shortest computation time. 
Regarding the limitations and the null model, the discussion becomes more intricate, acknowledging that there are no established guidelines, 
and the selection of a model also relies on the specific circumstances and characteristics of the actual network under investigation. 
The analysis in \cite{impactofnullmodelsecology} examined the statistical significance of several graph metrics using different null models and combined with an \gls{LRA} randomization approach.  
Within this framework, the entire null model space is explored and the authors argue that assessing the consistent significance of metrics, 
irrespective of the null model, as well as identifying those that are only occasionally significant, provides a more comprehensive perspective. \\

We use several null models and algorithms to construct random networks that adhere to a specific set of constraints $\mathbf{C}$. 
In the present study, we specifically examine undirected, directed, and bipartite graphs,
where the constraints are determined by their degree sequence(s). 
The detailed methodology is provided in Appendix \ref{appendix:methods}.

\subsection{A famous graph from the literature} 
\begin{figure}[p]
  \centering
  \includegraphics[width=0.99\textwidth]{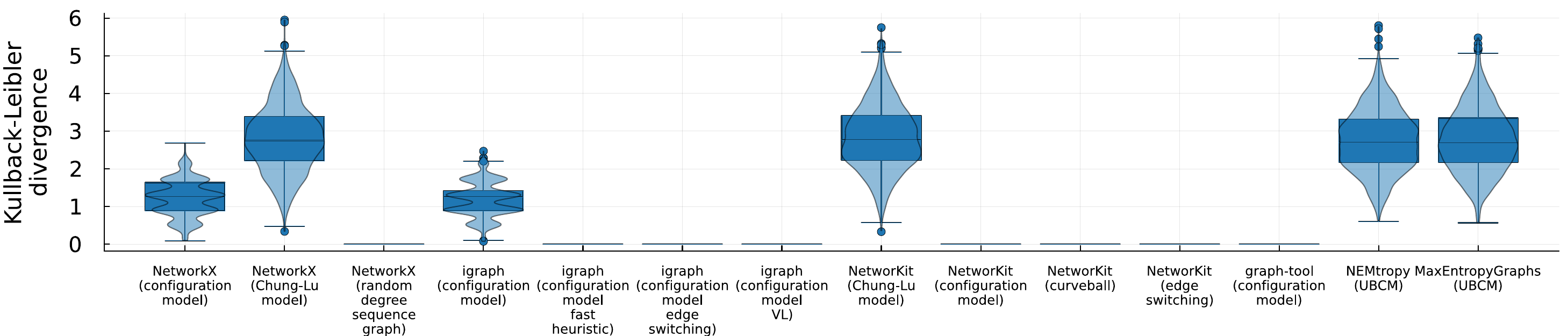}\\
  \centering{(a)}\\
  \includegraphics[width=0.99\textwidth]{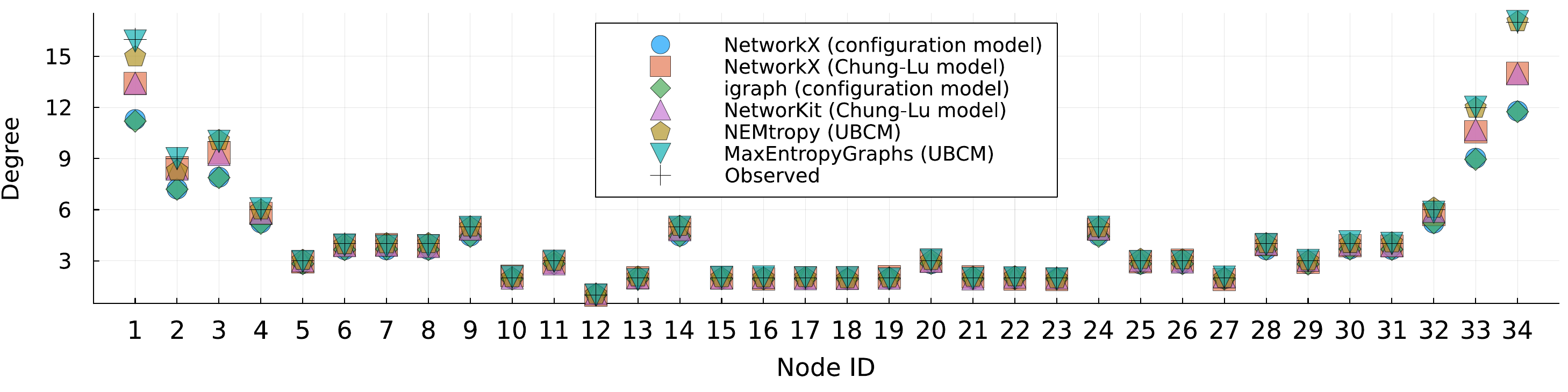}\\
  \centering{(b)}\\
  \includegraphics[width=0.99\textwidth]{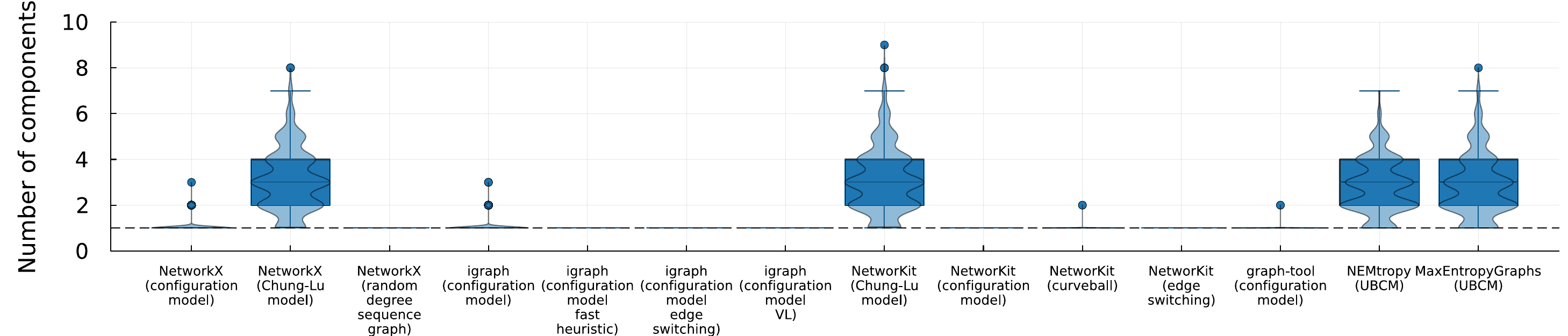}\\
  \centering{(c)}\\
  \includegraphics[width=0.99\textwidth]{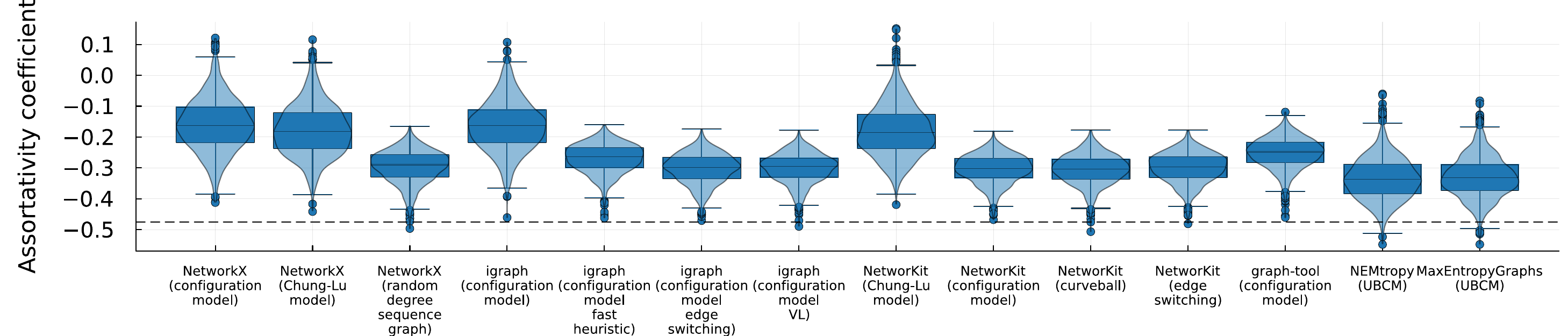}\\
  \centering{(d)}
  \caption[Karate Club network]{Karate Club network (1000 samples per model) (a) Kullback-Leibler divergence between the degree distribution of the observed graph and the randomised graphs. (b) Mean degree of each node of the randomised graphs. (c) Number of components in the randomised graphs. (d) Assortativity coefficient of the randomised graphs. The dashed lines indicate the values of the metric in the observed graph.}
  \label{chrand:fig:zachary}
  \Description{}
\end{figure}

First, we will analyze the well-known Zachary's karate club network, specifically the one that has 78 edges, as an illustrative example \cite{zachary1977information}.
We start by analyzing the level of adherence to imposed constraints, as shown in Figure \ref{chrand:fig:zachary}a. 
As expected, microcanonical models accurately reproduce the degree sequence, but canonical models do not. 
There are two obvious discrepancies in the \emph{configuration model} functions when using a stub-matching algorithm in NetworkX and igraph. 
These functions permit the presence of self-loops and multiple edges, resulting in a graph that does not exactly matches the observed degree sequence. 
This is noticeable when examining the average degree of each node in the randomized graphs, as depicted in Figure \ref{chrand:fig:zachary}b.
For the canonical models, we expect that the average degree of each node will be in close vicinity to the degree indicated in the degree sequence. 
The approaches of entropy maximisation successfully reproduce the degree sequence on average, while the Chung-Lu models from NetworkX and igraph are moderately successful. 
The discrepancy is especially noticeable in the case of nodes with a high degree.\\

After evaluating the degree to which the constraints are adhered to, we will now analyze various metrics and determine their statistical significance. 
Figure \ref{chrand:fig:zachary}c shows the distribution of the number of components in the random graphs. 
The majority of microcanonical models generate networks that are composed of a single component, 
this is partly due to the intentional design of certain algorithms. 
Conversely, the canonical models produce graphs that consist of several components in this particular scenario. 
The requirement for a singular component in the randomized graphs can impact the choice of the randomization method. 
The assortativity coefficient of the random graphs is shown in Figure \ref{chrand:fig:zachary}d. 
Although there may be some differences across the models, the overall findings are the same: 
the observed graph shows a greater level of disassortativity when compared to the random graph ensembles. \\

It is important to acknowledge the amount of time required to create the sample of random graphs. 
Among all the techniques used to produce a sample for this small graph, NetworkX, via its \texttt{random\_degree\_sequence\_graph} function, 
showed the longest duration, over 1.5 hours, when executed on a single core of an Apple M2 processor. 
On the other hand, the most efficient techniques were capable of generating a sample in under a minute.\\

\begin{figure}[p]
  \centering
  \includegraphics[width=0.99\textwidth]{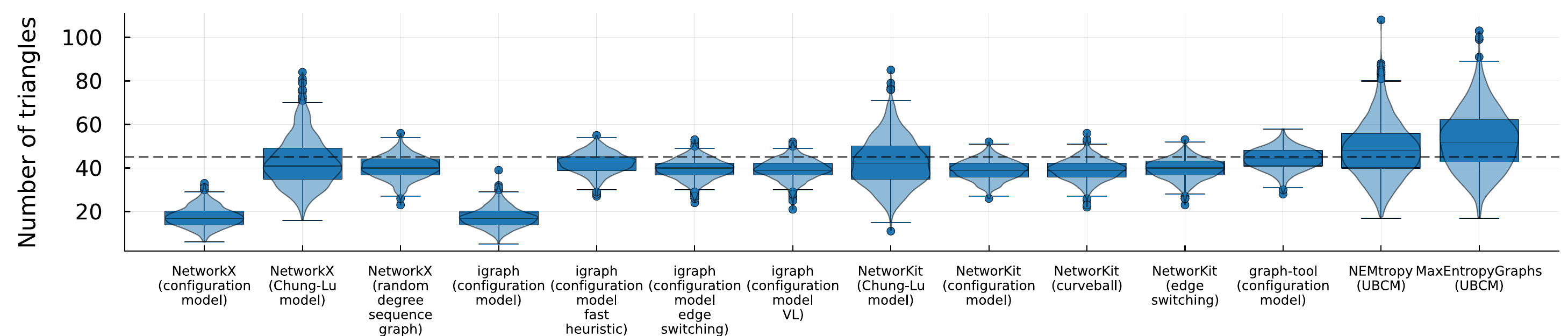}\\
  (a)\\
  \includegraphics[width=0.99\textwidth]{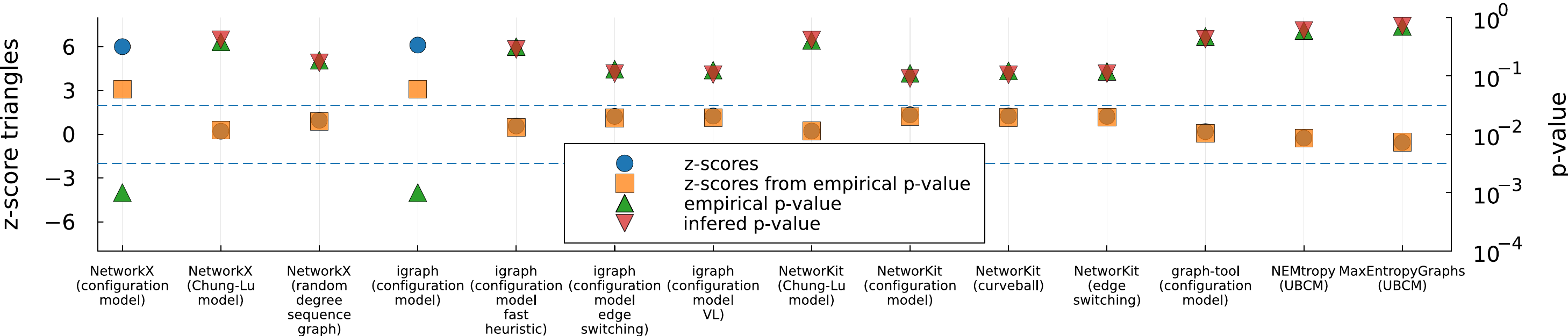}\\
  (b)
  \caption[Karate Club network (triangles)]{Karate Club network (1000 samples per model) (a) Distribution of the number of triangles. The dashed line indicates the number of triangles in the observed graph (b) Computed $z$-scores (left axis) and $p$-values (right axis) of the number of triangles. The dashed lines indicate the thresholds for statistical significance of the z-scores.}
  \label{chrand:fig:zachary:triangles}
  \Description{}
\end{figure}

With regard to network motifs, we may examine the distribution of the number of triangles in the graphs, as shown in Figure \ref{chrand:fig:zachary:triangles}a. 
When evaluating the statistical significance of the number of triangles, it becomes apparent that the conclusions vary across different methods and algorithms. 
The distributions of the number of triangles obtained from the samples generated by the \emph{configuration model} 
functions using stub-matching from NetworkX and igraph differ significantly from the distributions obtained with the other models. 
This can be attributed to the fact that the samples do not adhere to the degree sequence. 
Figure \ref{chrand:fig:zachary:triangles}b shows the computed $z$-scores and $p$-values for the number of triangles. 
The $z$-score and empirical $p$-values are calculated using the approach described in Section \ref{Ch:random:significance}. 
The $z$-score obtained from the empirical $p$-value is computed assuming a normal distribution. 
The estimated $p$-value is inferred from the $z$-score, assuming a normal distribution. 
The number of triangles, the $z$-score, and the $z$-score calculated from the empirical $p$-value are highly comparable. 
Nevertheless, this is not always the case. 

\subsection{An example from management science}
The authors of \cite{LIU2021437} analyzed a network of collaborative interactions among numerous contractors in the electronic database of China's National Quality Award Projects. 
A motif-based approach is used to depict the specific patterns of local relationships inside project networks and showcase the evolution of these relationship patterns.  
The largest network that was analysed had a total of 752 nodes and 16,631 edges. 
A \gls{LRA} approach was employed to create random graphs with a prescribed degree sequence. 
The statistical significance of various subgraphs was assessed using the $z$-score. 
Within the context of the project network, the various subgraphs are assigned an interpretation. 
For instance, a $4$-clique represents a group of four contractors who collaborate on the construction of at least one project. 
The authors argue that presence and structure of these subgraphs could be used to provide guidance to project organizations in selecting suitable 
governance systems for efficiently managing inter-organizational relationships.\\

The authors prioritize the significance of the subgraphs and analyze their counts, which are restricted to positive values. 
Nevertheless, for multiple subgraphs, the mean of the counts falls within one standard deviation of zero. 
This indicates that the distribution of subgraph counts is unlikely to follow a normal distribution, 
and therefore the $z$-score may not accurately measure statistical significance. 
Taking into account the empirical $p$-value could have resulted in different outcomes for some subgraphs. 
Additionally, the authors did not disclose information regarding the number of rewiring trials conducted. 
The relevance of this will become apparent in a later illustration. 
It is important to mention that in this specific application, due to the very high values recorded for certain subgraphs, 
the final results are likely to have remained mostly unchanged with this randomization decision. 
However, in other applications, this may not be true.

\subsection{A food web}
Food webs are used in biology to depict the trophic interactions among species within an ecosystem. 
Mapping out these interactions results in a directed graph. 
In this graph, the species are represented by nodes, and the feeding interactions are represented by edges, indicating the flow of energy from the prey to the predator. 
Within the context of food webs, the statistical significance of directed subgraphs can be of interest, and the prevalence 
of three-node subgraphs (cf. Figure \ref{fig:basics:directedsubgraphs}), or triads, is often used to describe and compare food webs \cite{Klaise:2017aa}. 
The significance profile of the triads has also been used to extract backbones of directed graphs \cite{Bai:2021aa}. 
We will demonstrate the differences between the randomization techniques for the Chesapeake Bay food web \cite{Chesapeakepaper}, 
which is a directed network consisting of $39$ nodes and $176$ edges.\\

\begin{figure}[p]
    \centering
    \includegraphics[width=0.95\textwidth]{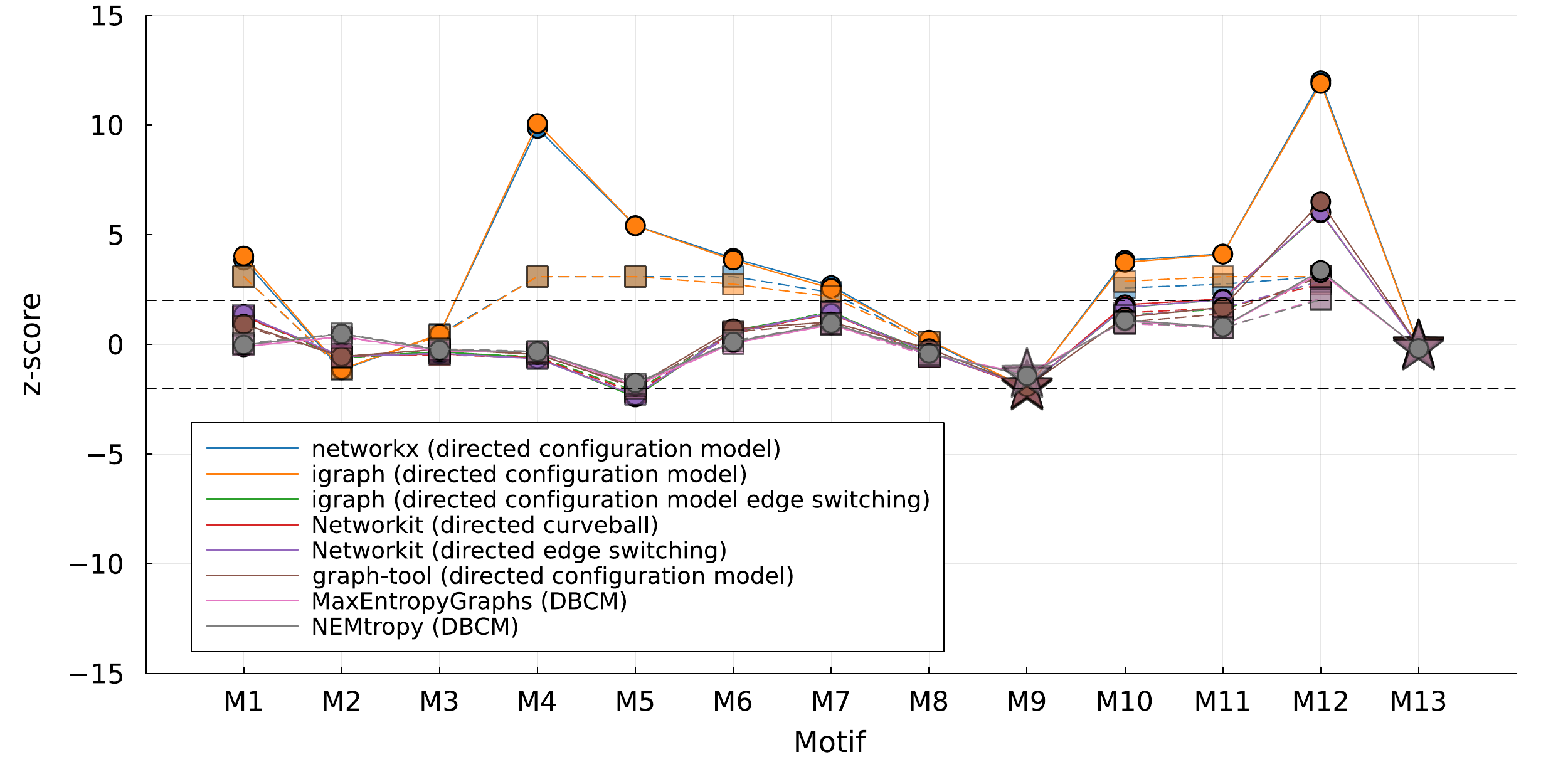}\\
    (a)\\
    \includegraphics[width=0.45\textwidth]{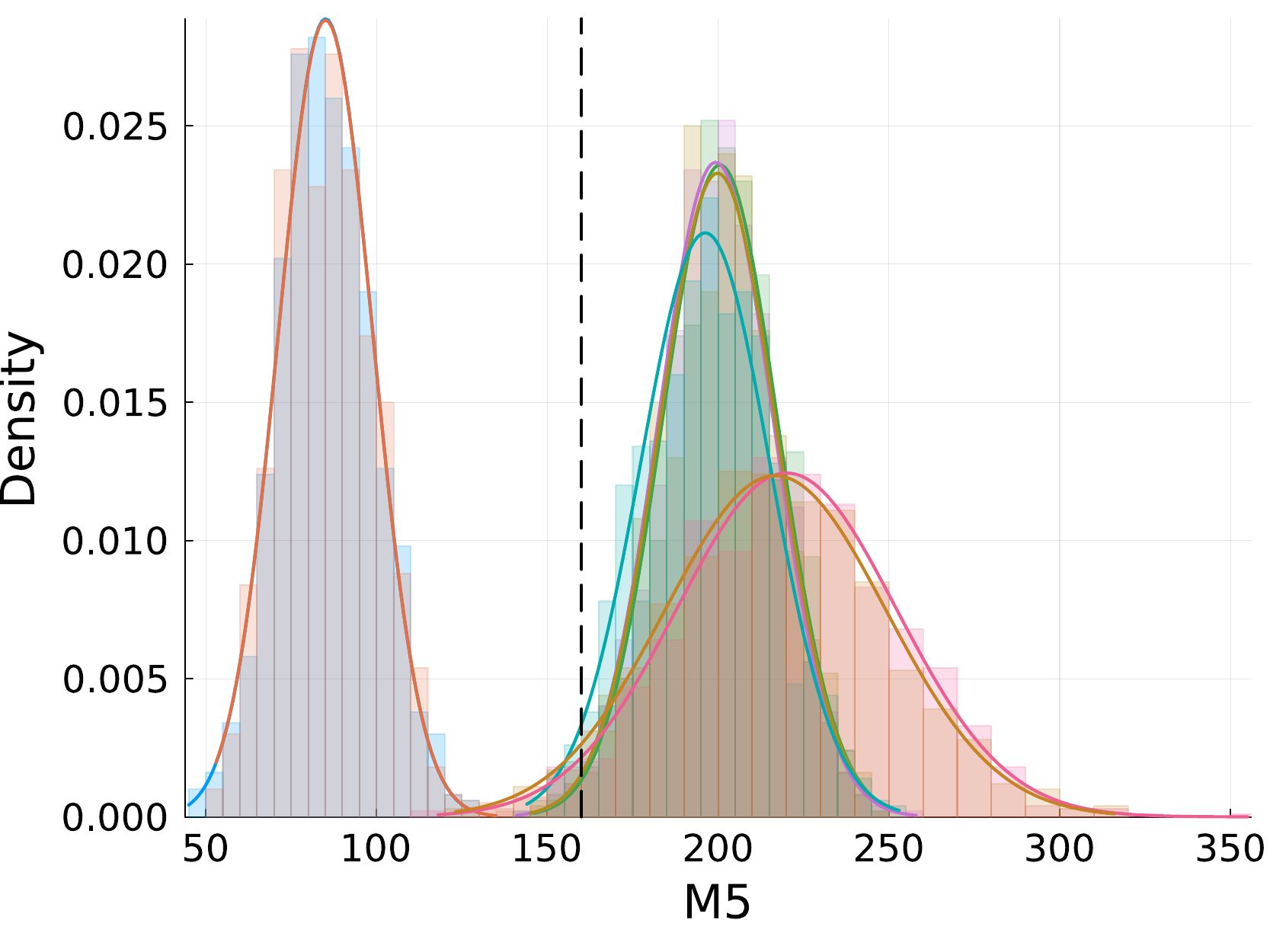}\includegraphics[width=0.45\textwidth]{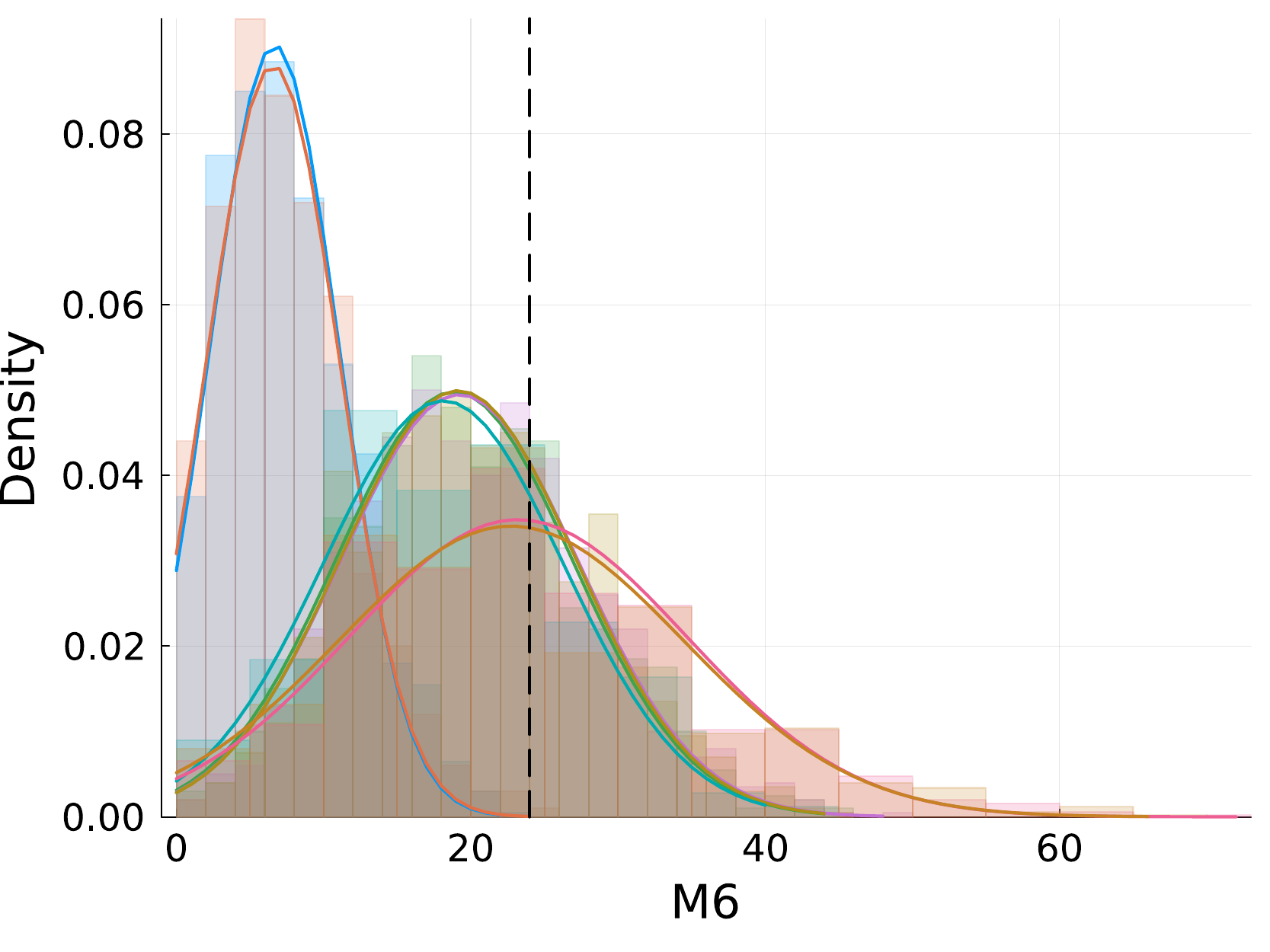}\\
    \includegraphics[width=0.55\textwidth]{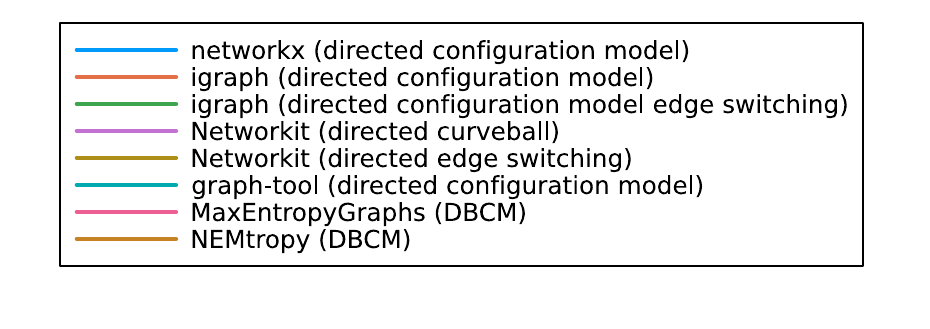}\\
    (b)
    \caption[Chesapeake Bay foodweb]{Chesapeake Bay foodweb (a) $z$-score of the different directed subgraphs involving $3$ nodes. The dashed horizontal indicates the threshold for statistical significance of the z-score. The circles indicate the $z$-score and the squares indicate the inferred $z$-score from the empirical $p$-value. Stars are used to indicate that the inferred $z$-score is not available, because the empirical $p$-value is $1$. (b) Distribution of the subgraph counts for $M5$ and $M6$. Both the histogram and the inferred normal distributions are shown. The dashed vertical lines indicates the observed values.}
    \label{chrand:fig:foodweb:significance}
    \Description{}
\end{figure}

Figure \ref{chrand:fig:foodweb:significance}a displays the $z$-score and 
the inferred $z$-score derived from the empirical $p$-value for various directed subgraphs consisting of $3$ nodes. 
The dashed horizontal line represents the threshold for statistical significance of the $z$-score, which is $1.96$ for a significance level of $0.05$. 
Several observations can be derived from this figure. 
First, it should be noted that the approaches employed by NetworkX and igraph, which rely on a configuration model with stub-matching, 
yield significantly different findings. 
As before, this can be explained by the presence of self-loops and multiple edges in the random graphs, which cause a violation of the degree sequence. 
Furthermore, the inferred $z$-scores for the same two techniques exhibit substantial differences relative to the $z$-score for some subgraphs. 
This can be explained by the deviation from normality in the distribution of the subgraph counts. 
Both $M9$ and $M13$ have an observed count of zero, 
which leads to the empirical $p$-value for the subgraphs $M9$ and $M13$ being $1$, resulting in an inferred $z$-score of $-\infty$. 
In the figure, these are denoted by a star marker placed at the corresponding $z$-score.\\

Figure \ref{chrand:fig:foodweb:significance}b displays the distribution of the subgraph counts for the subgraphs $M5$ and $M6$. 
This figure illustrates the variations in the number of subgraphs seen across various randomization methods. 
Additionally, it demonstrates the issues that can occur when assuming normality for subgraph counts with small values and, comparatively, a large variance. 
The subgraph counts for $M6$ obtained from the configuration model with stub-matching using NetworkX and igraph have a 
considerable percentage ($7.4\%$) of the inferred distribution that falls outside the feasible range.

\subsection{An example from Criminology} 
As a first example of a bipartite network, we will examine a network consisting of crimes and individuals who have been involved in at least one crime as a suspect, 
victim, witness, or both a suspect and victim simultaneously. 
An edge between two nodes $i$ and $\mu$ indicates that an individual $i$ was involved in a criminal activity $\mu$. 
The network consists of a total of 1,380 nodes, representing 829 individuals and 551 crimes, and 1,476 edges. 
The network data was obtained from \cite{konect}. 
A criminologist may find it relevant to examine the statistical importance of particular subgraphs, 
or the social network of criminals derived by projecting the bipartite network onto the layer of criminals, 
where two individuals are linked if they have jointly been involved in a crime. 
The various subgraphs present in this network possess meaningful real-world interpretations. 
For instance, a $\mathcal{V}$-motif can be understood as two individuals who participated in the same criminal activity, 
or as an individual who was involved in two separate crimes (depending on the perspective). 
Similarly, squares can be regarded as a pair of individuals who were jointly involved in two separate criminal activities. \\

\begin{figure}[p]
  \centering
  \includegraphics[width=0.875\textwidth]{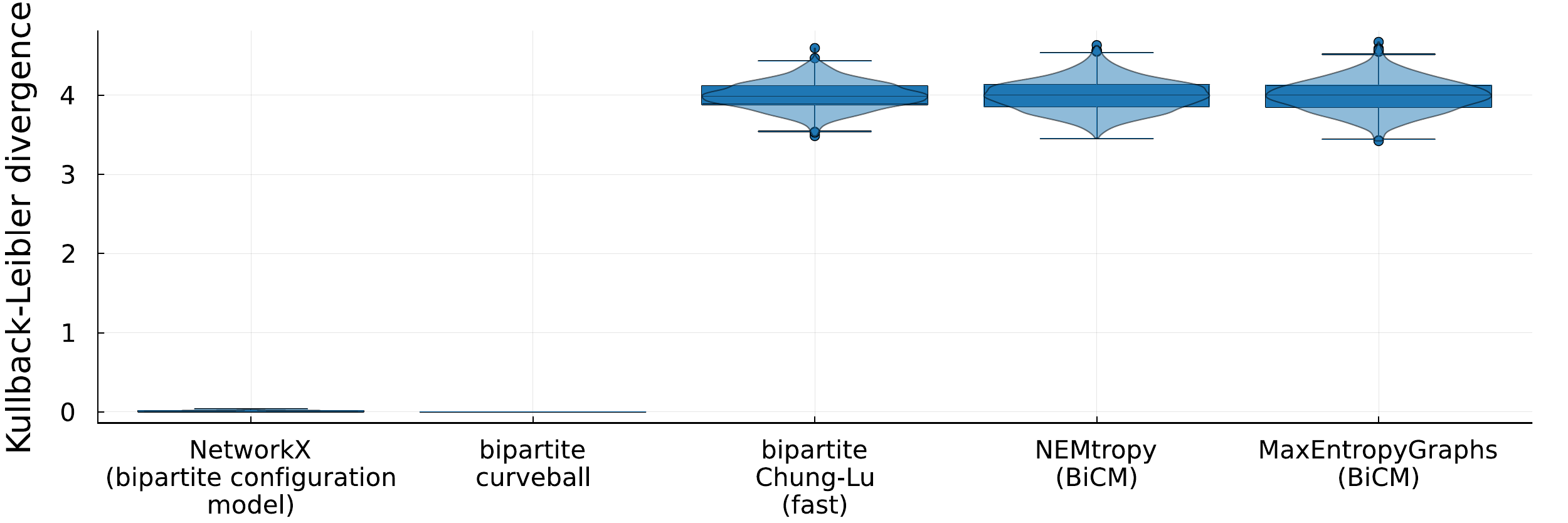}\\
  (a)\\
  \includegraphics[width=0.9\textwidth]{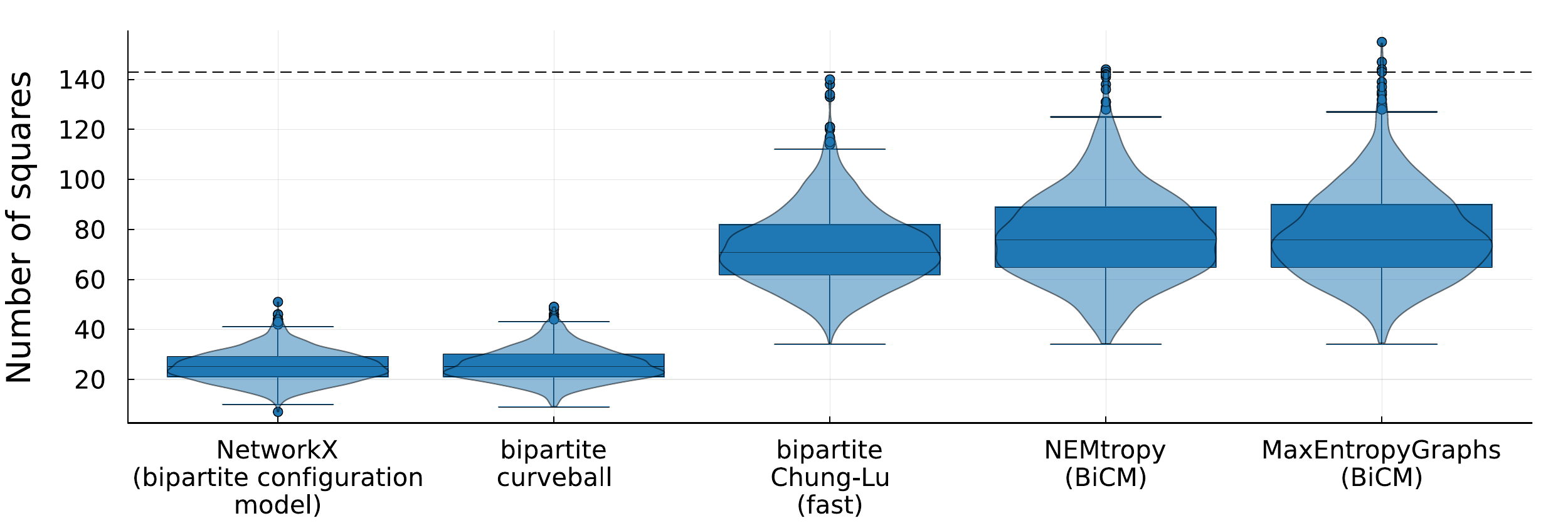}\\
  (b)\\
  \includegraphics[width=0.45\textwidth]{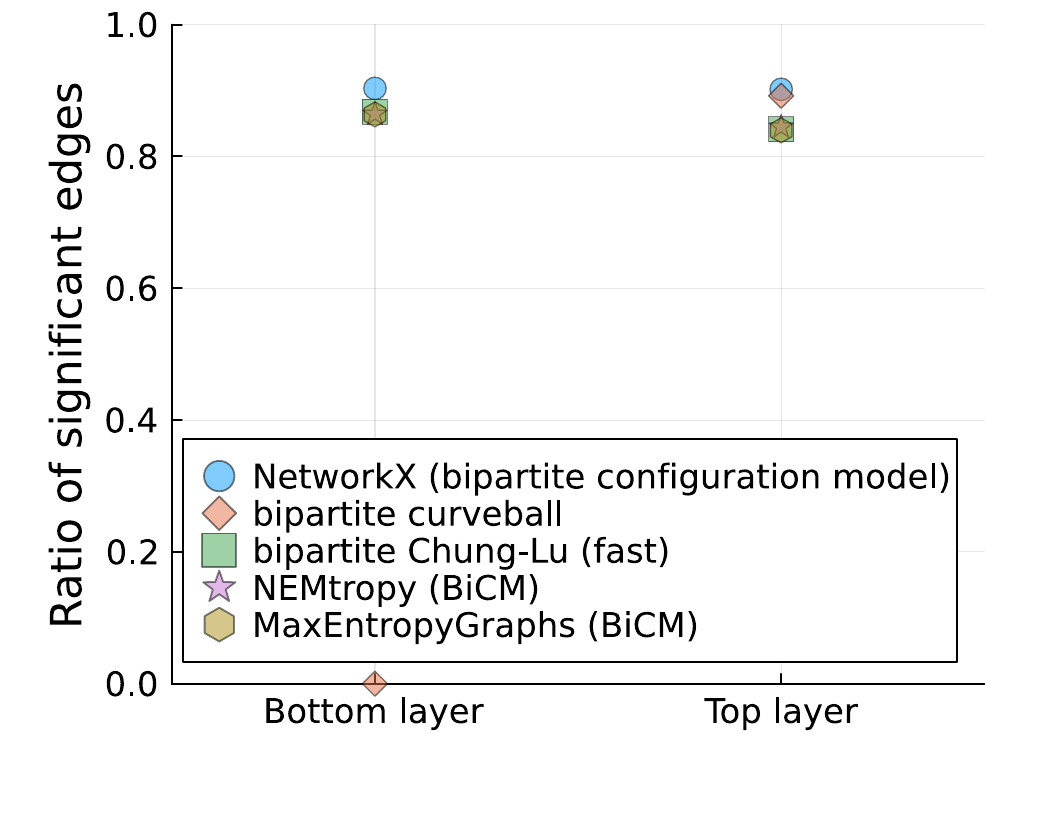}
  \hfill
  \includegraphics[width=0.45\textwidth]{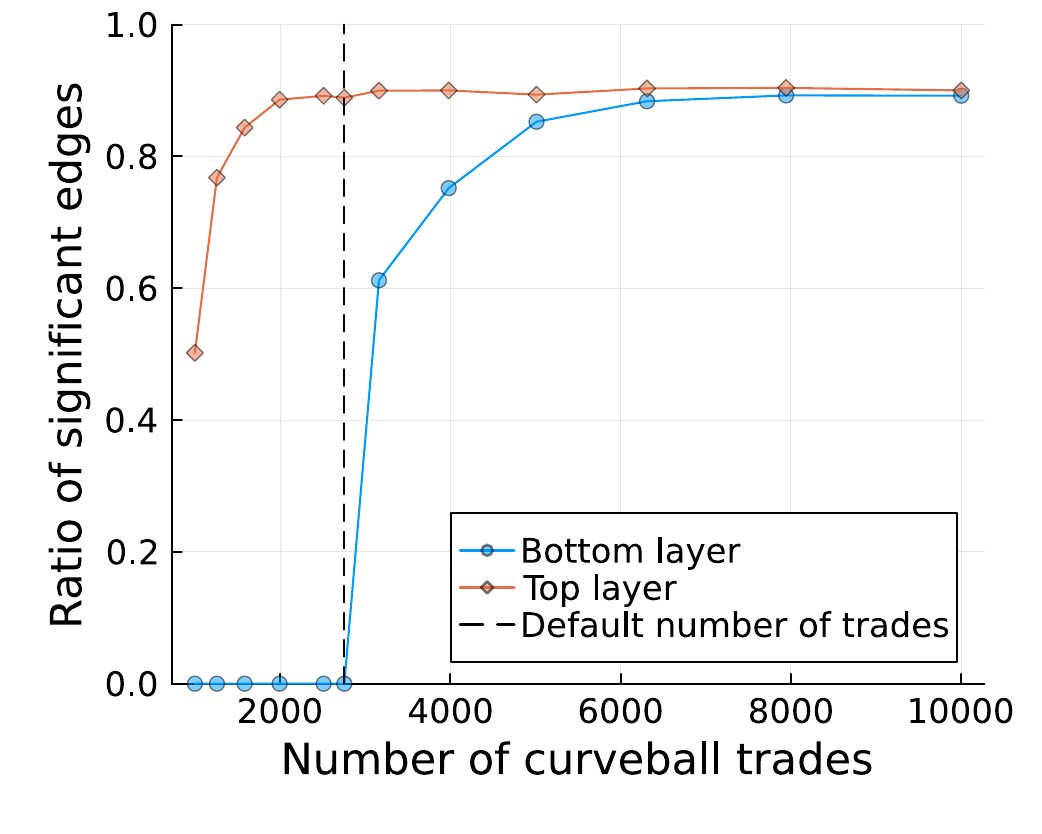}\\
  (c) \hspace{0.5 \textwidth} (d)\\
  \caption[Criminology network]{Criminology network (1000 samples per model) (a) Kullback-Leibler divergence between the degree distribution of the observed graph and the randomised graphs. (b) Distribution of the number of squares in the randomised graphs. The dashed horizontal line indicates the number of squares in the observed value. (c) Difference in fraction of statistically significant edges of the bipartite graph projected on either layer. (d) Evolution of the ratio of statistically significant edges for the curveball algorithm.}
  \label{chrand:fig:criminology}
  \Description{}
\end{figure}

The adherence to the constraints is illustrated in Figure \ref{chrand:fig:criminology}a. 
The curveball method is the sole algorithm that exactly replicates the degree sequence. 
The stub-matching algorithm from NetworkX experiences a reduction in its degree due to multiple edges being discarded. 
The canonical methods all show similar distribution of the Kullback-Leibler divergence.\\ 

Within the present context, the square subgraph illustrates a relationship between the individuals and the incidents. 
It could suggest a collaboration, recurring victimization, or participation as a witness, 
highlighting patterns of behavior or association between the individuals and the crimes.
Figure \ref{chrand:fig:criminology}b displays the distribution of the number of squares for the various randomization approaches. 
We can observe that the conclusions are consistent across the various models:  
The number of squares in the observed graph is higher than the number of squares in the randomised graphs. 
Nevertheless, there is a disparity in the distribution of the number of squares between the two conceptual approaches to randomization. \\

We will now consider the monopartite projections of the bipartite graph onto its layers.  
The fraction of the number of edges of the projections of the bipartite graph is shown in Figure \ref{chrand:fig:criminology}c. 
Somewhat surprisingly, the curveball method does not consider any of the edges in the bottom layer to be statistically significant.
This highlights a possible concern with the bipartite curveball technique, 
as well as with any methods that employ a Markov chain Monte Carlo approach. 
The number of rewiring tries, also referred to as trades, is a crucial parameter for these approaches. 
The results in Figure \ref{chrand:fig:criminology}c represents what was obtained by using the default value for generating random graphs, 
as described in the original publication of the Curveball method \cite{curveballmethod}. 
By increasing the number of transactions, we can find the equilibrium distribution of the Markov chain 
and arrive at similar results as the other models. 
The evolution of the fraction of statistically significant edges in function of the number of trades is shown in Figure \ref{chrand:fig:criminology}d.
Assuming a sufficiently large number of trades, the results within the conceptual approaches are coherent with one another, 
while there are differences in the number of statistically significant edges between the two conceptual approaches.\\

\subsection{An example from Sociology}
This case study uses a bipartite graph created from the documented sexual encounters involving 6,624 anonymous escorts 
and 10,106 persons who engaged in purchasing sexual services. The data for this graph was collected from an online community \cite{doi:10.1073/pnas.0914080107}. 
The original study's authors analyzed the network's evolution over time and observed the presence of preferred attachment behavior. 
In addition, when comparing the observed graph to a randomized graph ensemble that preserves the same degree sequence, 
negative degree correlations (disassortativity), and a higher number of 'four-cycles' (squares) than expected were observed.  
In this particular context, the density of squares can offer valuable insights regarding the spread of diseases.  
A later study did a thorough examination of the transmission of infections within the same network \cite{10.1371/journal.pcbi.1001109}. 
Upon replicating this study, we discovered similar results regarding the density of squares and the assortativity of degrees across all randomization models, 
which aligns with the findings of the original authors. \\

\begin{figure}[p]
  \centering
  \includegraphics[width=0.9\textwidth]{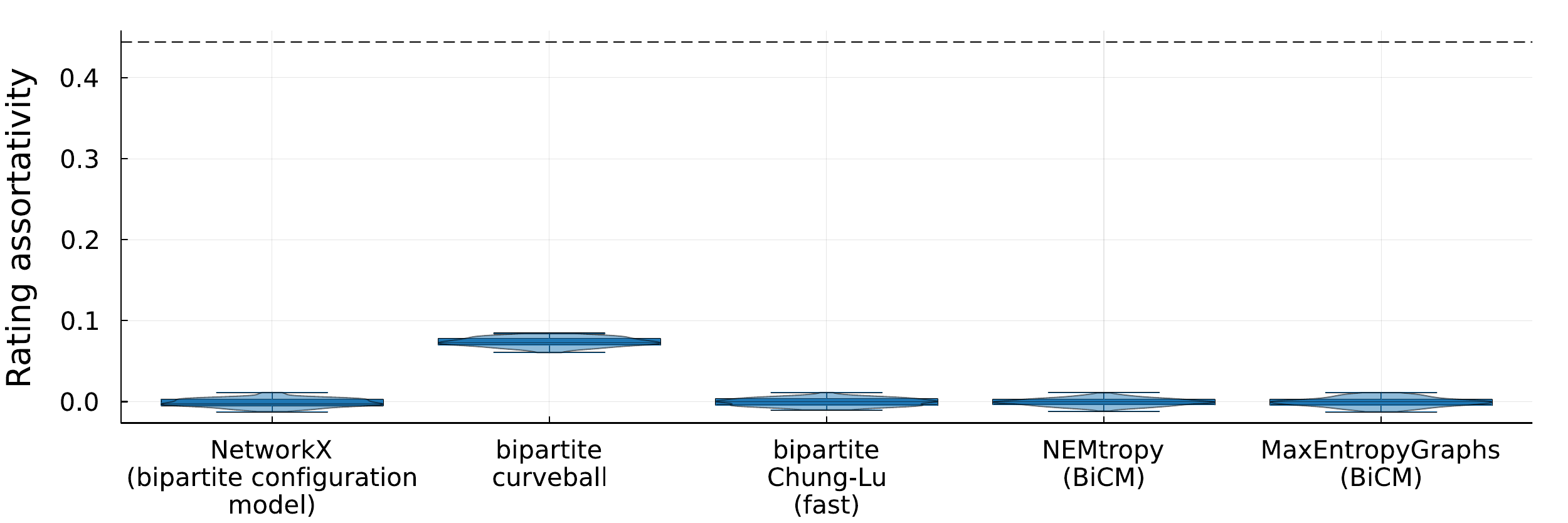}\\
  (a)\\
  \includegraphics[width=0.9\textwidth]{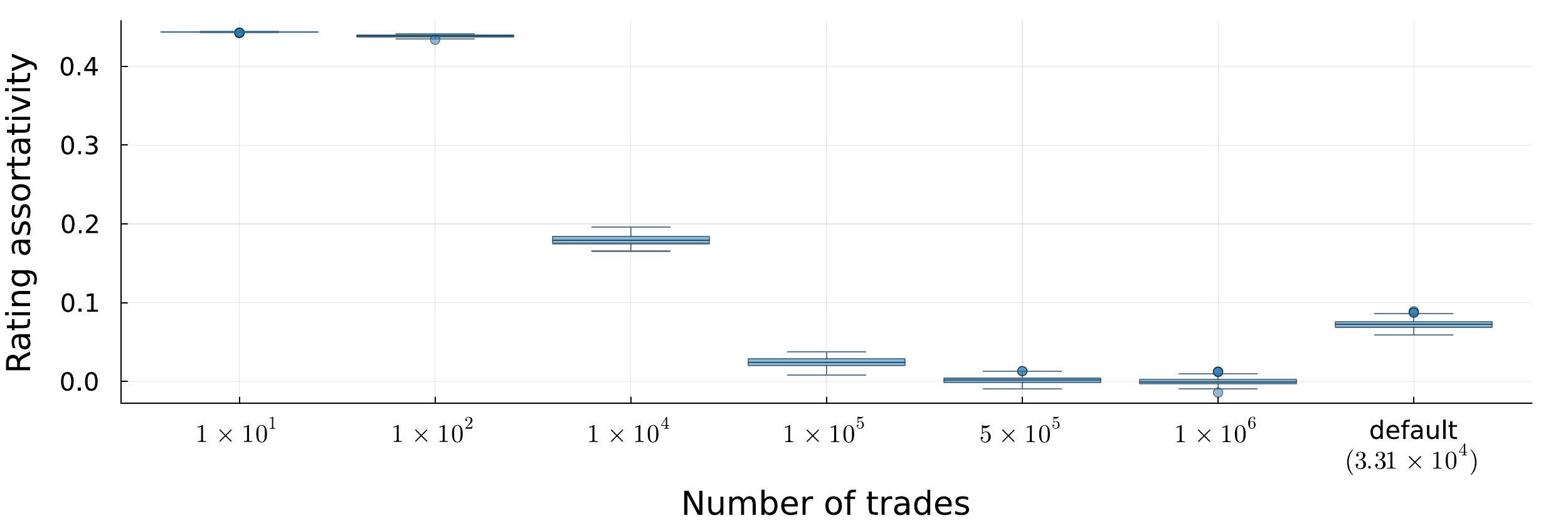}\\
  (b)
  \caption[Rating assortativity of the sex workers network]{Rating assortativity of the sex workers network  (100 samples per model) (a) Rating assortativity per model. (b) Evolution of the rating assortativity for an increasing number of trades for the curveball method (100 samples per trade).}
  \label{chrand:fig:escort:rating:assortativity}
  \Description{}
\end{figure}

To provide further insight, let us analyze the rating assortativity of the bipartite graph. 
Using the original data, we can calculate the mean rating provided by the buyer and the mean rating received by an escort. 
Afterwards, we may examine the assortativity of the bipartite graph using these ratings. 
This can reveal whether there is a propensity for buyers to frequently engage with escorts affiliated with a particular rating category. 
For example, are customers who give positive evaluations more inclined to interact with escorts who routinely receive positive ratings? 
The results of this analysis are shown in Figure \ref{chrand:fig:escort:rating:assortativity}a. 
All models concur that the observed graph is more assortative than the randomised graphs. 
The rating assortativity distribution obtained from the bipartite yields graphs that exhibit a minor assortativity. 
As was the case for the previous example, this is again due to an insufficient number of trades.
By increasing the number of trades, we could to find the equilibrium distribution of the Markov chain 
and arrived at the same result as the other models. 
The evolution of the rating assortativity in function of the number of trades is shown in Figure \ref{chrand:fig:escort:rating:assortativity}b.\\

\section{Best Practices and Recommendations}
Through comparative analysis of randomization methods for several graph types, we showed how similar conceptual methods, 
using a different algorithm, and different conceptual methods can lead to different outcomes. 
Thus underscoring the need to consider the nature of the observed graph when selecting a randomization method. 
Based on the observations made in Section \ref{sec:comparison},  
we provide a number of recommendations for researchers working with the types of random graphs that were discussed in this study.\\

It is important to consider the time required to generate random graphs. 
This can vary significantly between different methods, and can be a limiting factor considering different randomization methods. 
Additionally, the topology of a graph can also affect the time required to generate random graphs. 
Knowing that a different conceptual approach can lead to different outcomes, 
it is important to consider the nature of the observed graph when selecting a randomization method. 
If the degree sequence can only exactly appear in the random graphs, the microcanonical approach is the way to go. 
However we have also observed that the stub matching algorithm may result in degree distribution discrepancies, while it should not. 
This can particularly become a problem when there are a few nodes with a very high degree and a majority of nodes with a low degree. 
When using a Markov chain Monte Carlo method, such as the \gls{LRA} or the curveball method, 
it is important to evaluate if the number of iterations was sufficient to reach the equilibrium distribution, 
as not doing so can lead to misleading conclusions.
The canonical models on the other hand are more flexible in terms of the constraints they can adhere to, 
but they do not guarantee that the degree sequence is preserved on average. 
In many cases it is also the most computationally efficient method, 
these models have the advantage that they can be compared with one another using information theory principles to select the best one.
When working with networks coming from real world data, it is important to consider the possibility of measurement errors. 
Some even argue that this possibility should be the rule rather than the exception \cite{Peel:2022aa}. 
This is an argument in favor of the maximum entropy ensembles - and by extension all canonical methods - that was also made in \cite{squartini2017maximum-entropy}. \\

In terms of null model selection, there is no one-size-fits-all answer, the method of choice will depend on the specific problem context. 
This was also mentioned in previous studies \cite{Neal:2023gol}. 
One might also consider comparing conclusions across several null models \cite{impactofnullmodelsecology}.
Finally, we also advice caution when using the $z$-score to evaluate the significance of metrics, 
given the potential of non-normally distributed metrics.  
In particular for positive valued metrics for which the sample mean is close to zero in terms of the number of standard deviations of the sample.\\

In addition to being aware of these potential pitfalls, we recommend that authors using randomization methods 
are transparent in their reporting of the randomization procedures and results. 
This includes providing a detailed description of the randomization method used, inluding the algorithm used. 
If applicable the number of iterations should also be reported, as well as any other relevant algorithmic parameters.
This will allow for a better understanding of the results, increase the reproducibility, 
and facilitate the comparison of results across different studies.

\section{Future Directions and Open Challenges}
In this study we solely focussed on degree preserving randomization methods, because these are the ones 
that are often used in practice and are readily available in tools typically used by researchers. 
However, even when restricting the analysis to unweighted graphs, there are many other randomization methods that could be considered. 
These leads two a twofold observation. 
First of all, there is a need for more comprehensive coverage of randomisation algorithms by `standard' software tools. 
The best algorithm in the world is of no use if it is not easily available to the broader research community. 
Secondly, given the differences in outcomes between the different methods, 
extending this study to additional randomization methods could provide further insights into the strengths and weaknesses of each method.\\

The analysis of more complex network structures, such as multilayer, temporal, and higher-order networks, 
and the development and implemention of randomization methods that can handle these structures are also important areas for future research. 
While some progress has been made in this direction, there is still much work to be done. 
Just as is the case for the unweighted graphs, the availability of these methods in widely used software tools is crucial. 
For the analysis of multilayer networks, MuxViz \cite{10.1093/comnet/cnu038} is an established package, 
which has a randomisation method based on the configuration model.
It fixes the fixed degree sequences, while destroying intra-layer degree correlations, but not inter-layer ones. 
For the Julia programming language, the package MultiLayerGraphs.jl \cite{Moroni_Monticone_MultilayerGraphs_2022} is available, 
which has a randomisation method based on the configuration model combined with stub matching. 
In R, there is the multilayer.ergm package \cite{Chen_2021}. Similar to the ergm package for single layer networks,
it allows for the fitting and statistical analysis of multilayer exponential random graph models, but not for the explicit generation of random graphs.\\

For analyzing complex systems with higher order interactions, several tools have also been developed.
The XGI is a Python library which allows that offers the extension of the Chung Lu model to higher-order networks and 
the degree-corrected stochastic block model \cite{Landry_XGI_2023}. 
Another example is the Julia package HyperGraphs.jl \cite{10.1093/bioinformatics/btac347}, 
which also allows representing higher-order relationships, and has limited randomisation functionalities.
Nothwithstanding these initiatives, there still is a lot of room for development and testing, 
not only in terms of the algorithms themselves, but also in terms of scalability for larger networks.\\

Maximum entropy ensembles have been defined multilayer an higher order networks as well (see \cite{2016, bianconi2013, realstats}), 
but despite the existance of these models, standardized tools to generate random graphs from these models are still lacking. 
In the future, we hope to further extend our analysis to include these more complex network structures and their randomization methods, 
and contribute to the development of tools that can handle them.

\section{Conclusion}
The objective of this study was to compare the outcomes of different randomization methods an algorithms for different graph types. 
We focussed on the degree preserving randomization methods, as these are commonly used, and 
limited our analysis to tools that are readily available to the broader research community. 
We found that the choice of randomization method can have a significant impact on the results of network analysis. 
Different approaches have limitations, such as the inability to exactly reproduce the imposed constraints, 
or not guaranteeing that a sufficient number of iterations were performed.
Maximum entropy ensemble methods are in many cases more flexible, computationally efficient, 
account for measurement errors, and can be compared using information theory principles.
While the choice of randomization method should be based on the specific research question and the nature of the observed graph, 
in general, the entropy maximization approach is a good starting point, 
and its models increasingly being made available in easy to use and well-documented software packages. 
This work is just a small snapshot, as new algorithms - even for simple graphs - are still being published (e.g. \cite{mannion2024fast}), 
evaluating and comparing these algorithms and integrating them into widely used software tools, 
is a continuous process that can be benificial for the broader research community.

\begin{acks}
This work was supported by the DAP/19-03 project of the Royal Military Academy of Belgium. 
\end{acks}

\printbibliography

\appendix

\section{Case study methodology}
\label{appendix:methods}
The tools mentioned in the main text cover a variety of programming languages and are implemented in various manners. 
In order to compare the samples of random graphs generated by each tool and its corresponding model(s), we will proceed as follows. 
Initially, the graphs that are produced will be transformed into a standardized format, particularly the edgelist format. 
Certain models have a chance to produce graphs that contain self-loops or multiple edges. 
These elements will be removed when exporting the graphs. 
This can influence the compliance with specified constraints, but this is deliberate, 
as the graphs we are concerned with are simple in practice and should therefore be compared to simple graphs. 
This will also demonstrate the importance of selecting the appropriate randomization method for the given task. 
Next, we use the JuliaGraphs ecosystem as a common platform to compute the required metrics and topological characteristics. 
When examining statistical significance, we will evaluate both the $z$-scores and empirical $p$-values. 
To emphasize the differences between the randomization methods, we examine a range of graphs from various areas and of varying sizes. 
By default, we produce a set of 1000 random graphs for each approach, unless stated otherwise.  
Depending on the nature of the graph and the network it represents, we use different metrics to quantify the topological aspects of interest.\\

As the graph type becomes more complex, the available options for generating random graphs using the tools earlier dwindle (cf. Table \ref{tbl:chrand:comparison}). 
When working with bipartite graphs, we have only two options. 
One potential approach is to use the \texttt{configuration\_model} function from NetworkX, which employs a stub-matching algorithm. 
Another option is to use the entropy maximization techniques made available by NEMtropy and MaxEntropyGraphs.jl to sample the \gls{BiCM}. 
In order to conduct a thorough comparison, we also included the curveball technique outlined in the supplementary material of \cite{curveballmethod}, 
as well as the fast bipartite Chung-Lu method introduced in \cite{DBLP:journals/corr/AksoyKP16}. \\

We use the Kullback-Leibler divergence \cite{KLdivergence} to quantify to assess to what extent the constraints of every model are respected by the randomization methods. 
This metric is a measure of how one probability distribution diverges from a second, expected probability distribution.
If we use $Q$ to denote the probability distribution of the constraint(s) in the observed graph, 
$P$ to denote the probability distribution of the constraint(s) in a randomized graph and $\mathcal{X}$ for the sample space of the constraints, 
we can express the Kullback-Leibler divergence as follows:
\begin{equation}
    D_{KL}(P \Vert Q) = \sum_{x \in \mathcal{X}} P(x) \log \frac{P(x)}{Q(x)}
\end{equation}
By convention, one sometimes uses $0 \log \frac{0}{x} = 0\text{ }\forall x\in \mathbb{R}$ and $x \log \frac{x}{0} = \infty$ for $x > 0$ to deal with zero values in the distribution \cite{probdistances}. 
To prevent the occurrence of infinite values, we use Laplace smoothing on the probability distributions, which means that we add a small value $\epsilon$ to each value in the distribution. 
The value of $\epsilon$ is chosen to be at least an order of magnitude smaller than $1/N$, ensuring minimal disruption of the degree distribution.

\end{document}